\documentclass[10pt,twocolumn,letterpaper]{article}

\usepackage{cvpr}
\usepackage{times}
\usepackage{epsfig}
\usepackage{graphicx}
\usepackage{amsmath}
\usepackage{amssymb}
\usepackage{subcaption}
\usepackage{booktabs}
\usepackage{stackengine}
\usepackage{bm}
\usepackage[export]{adjustbox}
\usepackage{enumitem}
\usepackage{float}

\usepackage[pagebackref=true,breaklinks=true,letterpaper=true,colorlinks,bookmarks=false]{hyperref}

\cvprfinalcopy 

\def\cvprPaperID{4261} 

\ifcvprfinal\fi

\begin{document}

\title{DuDoNet: Dual Domain Network for CT Metal Artifact Reduction}

\author{Wei-An Lin*\textsuperscript{1} \hspace{2mm} Haofu Liao*\textsuperscript{2} \hspace{2mm} Cheng Peng\textsuperscript{1} \hspace{2mm} Xiaohang Sun\textsuperscript{3} \hspace{2mm} Jingdan Zhang\textsuperscript{4} \\Jiebo Luo\textsuperscript{2} \hspace{2mm} Rama Chellappa\textsuperscript{1} \hspace{2mm} Shaohua Kevin Zhou\textsuperscript{5,6}\\
\textsuperscript{1}University of Maryland, College Park \hspace{2mm}
\textsuperscript{2}University of Rochester \hspace{2mm}
\textsuperscript{3}Princeton University \\
\textsuperscript{4}Z2W Corporation \hspace{2mm}
\textsuperscript{5}Chinese Academy of Sciences \hspace{2mm}
\textsuperscript{6}Peng Cheng Laboratory, Shenzhen}

\maketitle
\thispagestyle{empty}

\begin{abstract}
{\let\thefootnote\relax\footnote{{* First two authors contributed equally.}}}
Computed tomography (CT) is an imaging modality widely used for medical diagnosis and treatment. CT images are often corrupted by undesirable artifacts when metallic implants are carried by patients, which creates the problem of metal artifact reduction (MAR). Existing methods for reducing the artifacts due to metallic implants are inadequate for two main reasons. First, metal artifacts are structured and non-local so that simple image domain enhancement approaches would not suffice. Second, the MAR approaches which attempt to reduce metal artifacts in the X-ray projection (sinogram) domain inevitably lead to severe secondary artifact due to sinogram inconsistency. To overcome these difficulties, we propose an end-to-end trainable Dual Domain Network (DuDoNet) to simultaneously restore sinogram consistency and enhance CT images. The linkage between the sinogram and image domains is a novel Radon inversion layer that allows the gradients to back-propagate from the image domain to the sinogram domain during training. Extensive experiments show that our method achieves significant improvements over other single domain MAR approaches. To the best of our knowledge, it is the first end-to-end dual-domain network for MAR.
\end{abstract}
\section{Introduction}

Computed tomography (CT) images reconstructed from X-ray projections allow effective medical diagnosis and treatment. However, due to increasingly common metallic implants, CT images are often adversely affected by metal artifacts which not only exhibit undesirable visual effects but also increase the possibility of false diagnosis. This creates the problem of metal artifact reduction (MAR), for which existing solutions are inadequate.

\begin{figure}[t!]
    \setlength{\abovecaptionskip}{3pt}
    \setlength{\tabcolsep}{2pt}
    \begin{tabular}[b]{cc}
        \begin{subfigure}[b]{.48\linewidth}
            \includegraphics[width=\textwidth,height=0.75\textwidth]{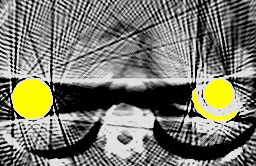}
            \caption{CT with metal artifacts}
            \label{fig:intro_ma}
        \end{subfigure} &
        \begin{subfigure}[b]{.48\linewidth}
            \includegraphics[width=\textwidth,height=0.75\textwidth]{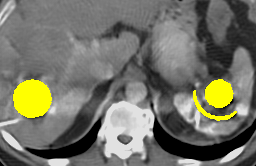}
            \caption{RDN\footnotemark~\cite{zhang-18-RDN}}
        \end{subfigure} \\
        \begin{subfigure}[b]{.48\linewidth}
            \includegraphics[width=\textwidth,height=0.75\textwidth]{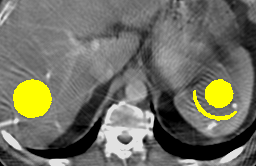}
            \caption{CNNMAR~\cite{Zhang-18-MAR-TMI}}
        \end{subfigure} &
        \begin{subfigure}[b]{.48\linewidth}
            \includegraphics[width=\textwidth,height=0.75\textwidth]{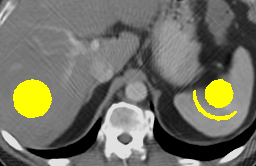}
            \caption{DuDoNet (Ours)}
            \label{fig:ours}
        \end{subfigure} \\
    \end{tabular}
    \caption{(a) Sample MAR results for a CT image with intense metal artifact. Metal implants are colored in yellow. (b) Artifacts are not fully reduced and a `white band' is present between the two implants. (c) Organ boundaries on the right are smeared out. (d) DuDoNet effectively reduces metal shadows and recovers fine details.}
\end{figure}

Unlike typical image restoration tasks such as super-resolution~\cite{ledig-17-srgan,zhang-18-RDN,wang-18-esrgan,zhong2018joint}, compression artifact removal~\cite{Zhang-18-dmcnn,Guo-16-dncnn}, and denoising~\cite{dabov-07-bm3d,makitalo-11-optimal,noise2noise}, metal artifacts are often \textit{structured and non-local} (e.g. streaking and shadowing artifacts as in Figure~\ref{fig:intro_ma}). Modeling such artifacts in image domain is extremely difficult. Therefore, before the emergence of deep learning, most existing works~\cite{Kalender-87-LI,duan2008metal,Meyer-10-nmar,Mehranian-13-L0} proposed to reduce metal artifact in the X-ray projection (sinogram) domain. The metal-corrupted regions are viewed as missing, and replaced by interpolated values. However, as the projections are taken from a single object under certain geometry, physical constraints should be satisfied by the enhanced sinogram. Otherwise, severe \textit{secondary artifacts} can be introduced in the reconstructed CT images.
\footnotetext{The residual dense network (RDN) proposed in~\cite{zhang-18-RDN} without up-scaling layers.}
\begin{figure*}[t]
    \centering
      \includegraphics[width=0.8\textwidth]{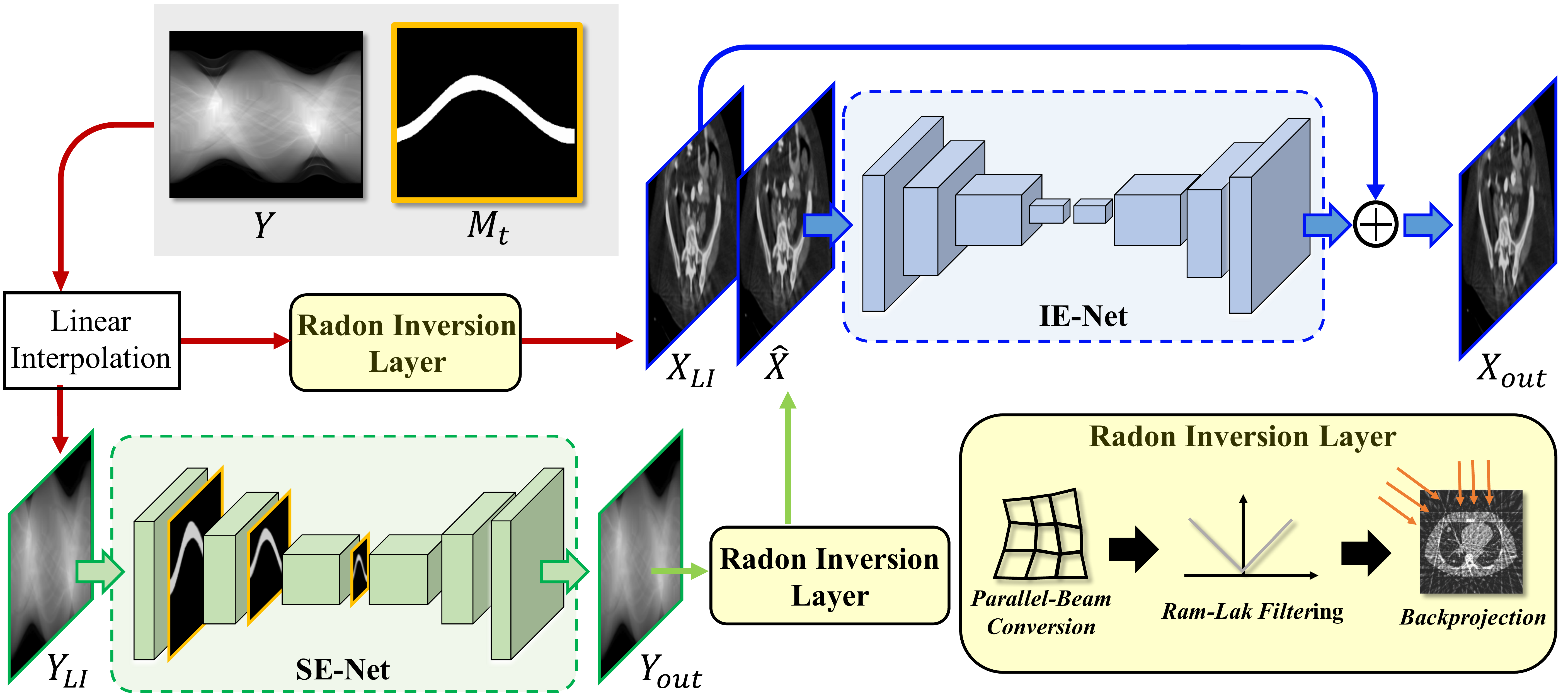}
    \caption{The proposed Dual Domain Network (DuDoNet) for MAR. Given a degraded sinogram $Y$ and a metal trace mask $\mathcal{M}_t$, DuDoNet reduces metal artifacts by simultaneously refining in the sinogram and image domains.}
    \label{fig:pipeline}
\end{figure*}

Recently, motivated by the success of deep learning in solving ill-posed inverse problems~\cite{zhang-18-RDN,wang-18-esrgan,noise2noise,pan-18-deblur,dehaze_zhang_2018,Ulyanov_2018_CVPR}, several works have been proposed to overcome the difficulties in MAR. Wang et al.~\cite{Wang-18-MAR-MICCAI} applied the pix2pix model~\cite{isola-17-pix2pix} to reduce metal artifact in the CT image domain. Zhang et al.~\cite{Zhang-18-MAR-TMI} proposed to first estimate a prior image by a convolutional neural network (CNN). Based on the prior image, metal-corrupted regions in the sinogram are filled with surrogate data through several post-processing steps for reduced secondary artifact. Park et al.~\cite{Park-17-sinogram} applied U-Net~\cite{ronneberger-15-unet} to directly restore metal-corrupted sinograms. Although metal artifacts can be reduced by these deep learning approaches, we will show that, despite the strong expressive power of deep neural networks, either image domain enhancement or sinogram domain enhancement is limited in being able to restore metal shadows and secondary artifact.




We hereby propose Dual Domain Network (DuDoNet) to address these problems by  \textit{learning two CNNs on dual domains to restore sinograms and CT images simultaneously}. Our intuition is that image domain enhancement can be improved by fusing information from the sinogram domain, and inconsistent sinograms can be corrected by the learning signal back-propagated from the image domain to reduce secondary artifacts. Specifically, we propose a novel network (Figure~\ref{fig:pipeline}) consisting of three parts: \textit{a sinogram enhancement network (SE-Net), a Radon inversion layer (RIL), and an image enhancement network (IE-Net)}. To address the issue that in the sinogram domain, information about small metal implants tends to vanish in higher layers of the network due to down-sampling, we propose a mask pyramid U-Net architecture for SE-Net, which retains metal mask information across multiple scales. The key to our dual-domain learning is RIL that reconstructs CT images using the filtered back-projection (FBP) algorithm and efficiently back-propagates gradients from the image domain to the sinogram domain. Based on RIL, we introduce a Radon consistency loss to penalize secondary artifacts in the image domain. Finally, IE-Net refines CT images via residual learning. Extensive experiments on CT images from hundreds of patients demonstrate that dual domain enhancement generates superior artifact-reduced CT images. 

In summary, we make the following contributions:
\begin{itemize}[noitemsep,topsep=0pt]
    \item We propose an end-to-end trainable dual-domain refinement network for MAR. The network is able to recover details corrupted by metal artifacts.
    \item We propose a mask pyramid (MP) U-Net to improve sinogram refinement. The MP architecture improves performance especially when small metallic implants are dominated by the non-metal regions.
    \item We propose a Radon inversion layer (RIL) to enable efficient end-to-end dual domain learning. RIL can benefit the community through its ubiquitous use in various reconstruction algorithms~\cite{Wurfl-16-MICCAI,Jin-17-TIP,Adler-18-TMI,Zhang-18-TMI}. 
    \item We propose a Radon consistency (RC) loss to penalize secondary artifacts in the image domain. Gradients of the loss in the image domain are back-propagated through RIL to the sinogram domain for improved consistency.
\end{itemize}

\section{Backgrounds and Related Works}
Tissues inside the human body such as bones and muscles have different X-ray attenuation coefficients $\mu$. If we consider a 2D slice of human body, the distribution of the attenuation coefficients $X = \mu(x,y)$ represents the underlying anatomical structure. The principle of CT imaging is based on the fundamental Fourier Slice Theorem, which guarantees that the 2D function $X$ can be reconstructed solely from its dense 1D projections. In CT imaging, projections of the anatomical structure $X$ are inferred by the emitted and received X-ray intensities through the Lambert-Beer Law~\cite{Beer}. We consider the following CT model under a polychromatic X-ray source with energy distribution $\eta(E)$:
\begin{align}
    \label{eq:sinogram}
    Y &= -\log \int \eta(E) \exp \left\{- \mathcal{P} X(E) \right\}dE, 
\end{align}
where $\mathcal{P}$ is the projection generation process, and $Y$ represents the projection data (sinogram). The 2D $X(E)$ is the anatomical structure (CT image) we want to recover from the measured projection data $Y$ .

For normal body tissues, $X(E)$ is almost constant with respect to the X-ray energy $E$. If we let $X = X(E)$, then
\begin{equation}
    \label{eq:approx}
    Y = \mathcal{P}X.
\end{equation}
Therefore, given measured projection data $Y$, the CT image $\hat{X}$ can be inferred by using a reconstruction algorithm $\mathcal{P^{\dagger}}$\footnote{We use $\mathcal{P^{\dagger}}$ to denote the linear operation for reconstruction.}: $\hat{X} = \mathcal{P^{\dagger}} Y$~\cite{kak}.

However, when metallic implants $I_M(E)$ are present, $X(E) = X + I_M(E)$, where $X(E)$ has large variations with respect to $E$ due to $I_M$. Eq. \eqref{eq:sinogram} becomes
\begin{equation}
\label{eq:sinogram_domain}
    Y = \mathcal{P}X -  \log \int \eta(E) \exp \{-\mathcal{P} I_M(E) \} dE,
\end{equation}
where the region of $\mathcal{P} I_M$ in $Y$ will be referred to as \emph{metal trace} in the rest of the paper.
When the reconstruction algorithm $\mathcal{P^{\dagger}}$ is applied,
\begin{equation}
\label{eq:image_domain}
    \mathcal{P^{\dagger}} Y = \hat{X} - \mathcal{P^{\dagger}} \log \int \eta(E) \exp \{-\mathcal{P} I_M(E) \} dE.
\end{equation}
The term after $\hat{X}$ in~\eqref{eq:image_domain} is the metal artifact.
It is clear that perfect MAR can be achieved only if the last term in Eq.~\eqref{eq:image_domain} is suppressed while the term $\hat{X}$ is unaffected. However, it is generally an ill-posed problem since both terms contribute to the region of metal trace.

\subsection{Inpainting-based Methods}
One commonly adopted strategy in MAR is to formulate sinogram completion as an image inpainting task. Data within the metal trace are viewed as missing and filled through interpolation. Linear interpolation (LI)~\cite{Kalender-87-LI} is a widely used method in MAR due to its simplicity. Meyer et al.~\cite{Meyer-10-nmar} proposed the NMAR algorithm, where sinograms are normalized by tissue priors before performing LI. NMAR requires proper tissue segmentation in the image domain, which is unreliable when severe metal artifacts are present. Mehranian et al.~\cite{Mehranian-13-L0} restored sinograms by enforcing sparsity constraints in the wavelet domain. In general, inpainting-based approaches fail to replace the data of $\mathcal{P} X$ in~\eqref{eq:sinogram_domain} within metal trace by consistent values. \emph{It is this introduced inconsistency in sinogram data that leads to noticeable secondary artifacts after reconstruction.}

\subsection{MAR by Iterative Reconstruction}
In iterative reconstruction, MAR can be formulated as the following optimization problem:
\begin{equation}
    \label{eq:recon}
    \hat{X} = \min_{X} \lVert (1 - \mathcal{M}_t) \odot (\mathcal{P} X - Y) \rVert^2 + \lambda R(X),
\end{equation}
where $\mathcal{M}_t$ is the metal trace mask. $\mathcal{M}_t=1$ on the metal trace and $\mathcal{M}_t=0$ otherwise. $R$ is some regularization function, e.g. total variation (TV)~\cite{Zhang-16-TV} and sparsity constraints in the wavelet domain~\cite{Zhang-18-JSR}. Eq.~\eqref{eq:recon} is often solved through iterative approaches such as the split Bregman algorithm. Iterative reconstruction usually suffers from long processing time as they require multiplying and inverting huge matrices in each iteration. More importantly, hand-crafted regularization $R(X)$ does not capture the structure of metal artifacts and would result in an over-smoothed reconstruction. Recently, Zhang et al.~\cite{Zhang-18-JSR} proposed a re-weighted JSR method which combines NMAR into~\eqref{eq:recon} and jointly solves for $X$ and interpolated sinogram. Similar to NMAR, the weighting strategy in re-weighted JSR requires tissue segmentation. In phantom study, better performance against NMAR is achieved by re-weighted JSR. However, the improvements remain limited for non-phantom CT images.

\subsection{Deep Learning based Methods for MAR}
Convolutional neural networks have the ability to model complex structures within data. Motivated by the success of DNNs in solving inverse problems, Gjesteby et al.~\cite{Gjesteby-17-sinogram} and Park et al.~\cite{Park-17-sinogram} proposed to refine sinograms using a CNN for improved consistency. Zhang et al.~\cite{Zhang-18-MAR-TMI} proposed a CNNMAR model to first estimate a prior image by a CNN and then correct sinogram similar to NMAR. However, even with the strong expressive power of CNNs, these approaches still suffer from secondary artifacts due to inconsistent sinograms.

Gjesteby et al.~\cite{Gjesteby-17-image}, Xu et al.~\cite{Xu-18-MAR} and Wang et al.~\cite{Wang-18-MAR-MICCAI} proposed to reduce metal artifact directly in the CT image domain. The metal artifacts considered in these works are mild and thus can be effectively reduced by a CNN. We will show in our experiments that image domain enhancement is not sufficient for mitigating intense metal shadows.
\section{Proposed Method}
As shown in Figure~\ref{fig:pipeline}, our proposed model consists of three parts: (a) a sinogram enhancement network (SE-Net), (b) a Radon inversion layer (RIL), and (c) an image enhancement network (IE-Net). Inputs to the model include a degraded sinogram $Y \in \mathbb{R}^{H_s \times W_s}$ and the corresponding metal trace mask $\mathcal{M}_t \in \{0,1\}^{H_s \times W_s}$. Notice that we use $H_s$ to represent the detector size and $W_s$ to represent the number of projection views. The region where $\mathcal{M}_t = 1$ is the metal trace. Given the inputs, we first apply LI~\cite{Kalender-87-LI} to generate an initial estimate for the sinogram data within metal trace. The resulting interpolated sinogram is denoted by $Y_{LI}$. SE-Net then restores $Y_{LI}$ within the metal trace through a mask pyramid U-Net architecture. To maintain sinogram consistency, we introduce a Radon consistency (RC) loss. A sinogram will be penalized by the RC loss if it leads to secondary artifacts in the image domain after passing through RIL. Finally, the reconstructed CT image $\hat{X} \in \mathbb{R}^{H_c \times W_c}$ is refined by IE-Net via residual learning. 

\subsection{Sinogram Enhancement Network}
Sinogram enhancement is extremely challenging since geometric consistency should be retained to prevent secondary artifact in the reconstructed CT image, so prior works only replace data within the metal trace. Similarly, given a metal-corrupted sinogram $Y$ and metal trace mask $\mathcal{M}_t$, SE-Net $\mathcal{G}_s$ learns to restore the region of $Y_{LI}$ in $\mathcal{M}_t=1$. In sinogram domain enhancement, when the metal size is small, or equivalently, the metal trace is small, information about metal trace is dominated by non-metal regions in higher layers of network due to down-sampling. To retain the mask information, we propose to fuse $\mathcal{M}_t$ through a mask pyramid U-Net architecture. The output of SE-Net is written as
\begin{equation}
    Y_{out} = \mathcal{M}_t \odot \mathcal{G}_s(Y_{LI}, \mathcal{M}_t) + (1 - \mathcal{M}_t) \odot Y_{LI}.
\end{equation}
We use an $L_1$ loss to train SE-Net:
\begin{align}
    \mathcal{L}_{\mathcal{G}_s} &= \lVert Y_{out} - Y_{gt} \rVert_1,
\end{align}
where $Y_{gt}$ is the ground truth sinogram without metal artifact.

\subsection{Radon Inversion Layer}
Although sinogram inconsistency is reduced by SE-Net, there is no existing mechanism to penalize secondary artifacts in the image domain. The missing key element is an efficient and \emph{differentiable} reconstruction layer. Therefore, we propose a novel RIL $f_R$ to reconstruct CT images from sinograms and at the same time allow back-propagation of gradients. We hightlight that trivially inverting $\mathcal{P}$ in existing deep learning frameworks would require a time and space complexity of $\mathcal{O}(H_s W_s H_c W_c)$, which is prohibitive due to limited GPU memory.

In this work, we consider the projection process $\mathcal{P}$ as the Radon transform under fan-beam geometry with arc detectors~\cite{kak}. The distance between an X-ray source and its rotation center is $D$. The resulting fan-beam sinograms $Y_{fan}$ are represented in coordinates $(\gamma, \beta)$. To reconstruct CT images from $Y_{fan}(\gamma, \beta)$, we adopt the fan-beam filtered back-projection (FBP) algorithm as the forward operation of RIL. 

Our RIL consists of three modules: (a) a parallel-beam conversion module, (b) a filtering module, and (c) a backprojection module. The parallel-beam conversion module transforms $Y_{fan}(\gamma, \beta)$ to its parallel-beam counterpart $Y_{para}(t, \theta)$ through a change of variables. The FBP algorithm in coordinate $(t, \theta)$ becomes more effective and memory-efficient than in $(\gamma, \beta)$. Parallel-beam FBP is then realized by the subsequent filtering and back-projection modules.

\textbf{Parallel-beam Conversion Module.} We utilize the property that a fan beam sinogram $Y_{fan}(\gamma, \beta)$ can be converted to its parallel beam counterpart $Y_{para}(t, \theta)$ through the following change of variables~\cite{kak}:
\begin{align}
    \begin{cases} 
        t = D \sin \gamma, \\
        \theta = \beta + \gamma.
    \end{cases}\label{eq:fan2para}
\end{align}
The change of variable in~\eqref{eq:fan2para} is implemented by grid sampling in $(t, \theta)$, which allows back-propagation of gradients. With $Y_{para}$, CT images can be reconstructed through the following Ram-Lak filtering and back-projection modules.
  
\textbf{Ram-Lak Filtering Module.} We apply the Ram-Lak filtering to $Y_{para}$ in the Fourier domain.
\begin{equation}
    Q(t, \theta) = \mathcal{F}^{-1}_t \left\{ |\omega| \cdot \mathcal{F}_t\left\{Y_{para}(t, \theta)\right\}  \right\},
\end{equation}
where $\mathcal{F}_t$ and $\mathcal{F}_t^{-1}$ are the Discrete Fourier Transform (DFT) and inverse Discrete Fourier Transform (iDFT) with respect to the detector dimension.  

\textbf{Backprojection Module.}
The filtered parallel-beam sinogram $Q$ is back-projected to the image domain for every projection angle $\theta$ by the following formula:
\begin{equation}
    X(u, v) = \int_0^\pi Q(u \cos \theta + v \sin \theta, \theta) d \theta.
    \label{eq:bp}
\end{equation}
It is clear from~\eqref{eq:bp} that the computation is highly parallel. We make a remark here regarding the property of RIL $f_R$. Due to the back-projection nature of $f_R$, the derivative with respect to the input $Y_{out}$ is actually the projection operation $\mathcal{P}$. That is, any loss in the image domain will be aggregated and projected to the sinogram domain. This desirable property enables joint learning in sinogram and image domains.

\begin{figure*}[t]
    \setlength{\abovecaptionskip}{3pt}
    \setlength{\tabcolsep}{2pt}
    \begin{tabular}[b]{c | c}
        \hspace{0.5cm}
        \begin{tabular}[b]{c}
        \begin{subfigure}[b]{0.2\linewidth}
            \setlength{\abovecaptionskip}{2pt}
            \includegraphics[width=\textwidth,height=0.8\textwidth]{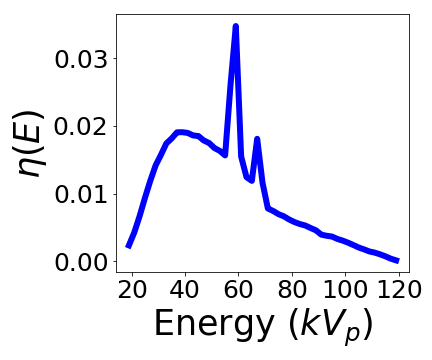}
        \end{subfigure} \\
        \begin{subfigure}[b]{.18\linewidth}
            \setlength{\abovecaptionskip}{2pt}
            \includegraphics[width=\textwidth,height=0.75\textwidth]{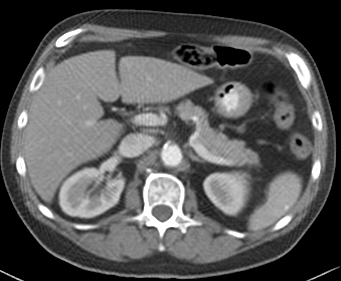}
        \caption*{Ground truth CT image}
        \end{subfigure}
        \end{tabular}
        &
        \setlength{\tabcolsep}{1pt}
        \begin{tabular}[b]{cccc}
        \begin{subfigure}[b]{.18\linewidth}
        \setlength{\abovecaptionskip}{2pt}
            \includegraphics[width=\textwidth,height=0.9\textwidth]{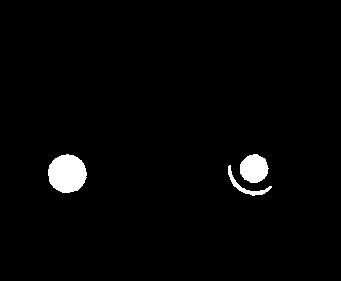}
        \end{subfigure} &
        \begin{subfigure}[b]{.18\linewidth}
            \setlength{\abovecaptionskip}{2pt}
            \includegraphics[width=\textwidth,height=0.9\textwidth]{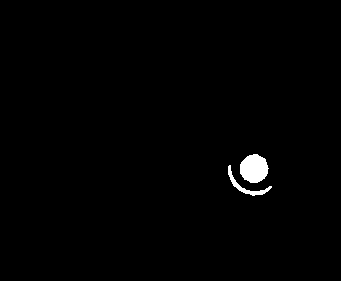}
        \end{subfigure} &
        \begin{subfigure}[b]{.18\linewidth}
            \setlength{\abovecaptionskip}{2pt}
            \includegraphics[width=\textwidth,height=0.9\textwidth]{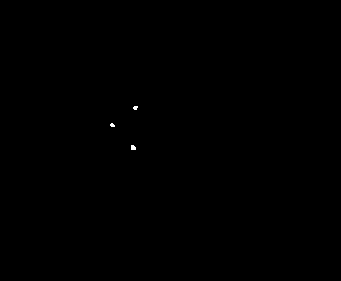}
        \end{subfigure} &
        \begin{subfigure}[b]{.18\linewidth}
            \setlength{\abovecaptionskip}{2pt}
            \includegraphics[width=\textwidth,height=0.9\textwidth]{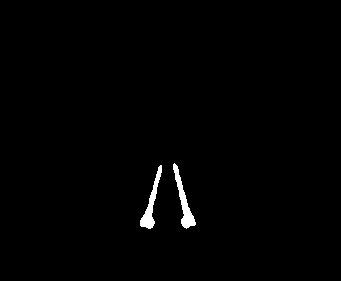}
        \end{subfigure}\\
        \begin{subfigure}[b]{.18\linewidth}
            \setlength{\abovecaptionskip}{2pt}
            \includegraphics[width=\textwidth,height=0.9\textwidth]{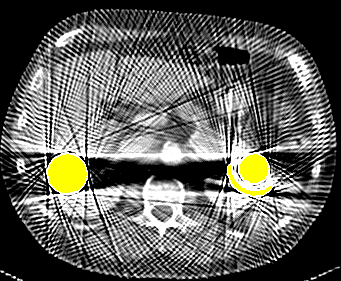}
        \end{subfigure} &
        \begin{subfigure}[b]{.18\linewidth}
            \setlength{\abovecaptionskip}{2pt}
            \includegraphics[width=\textwidth,height=0.9\textwidth]{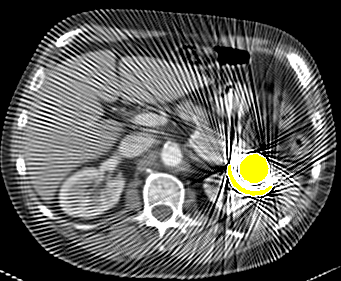}
        \end{subfigure} &
        \begin{subfigure}[b]{.18\linewidth}
            \setlength{\abovecaptionskip}{2pt}
            \includegraphics[width=\textwidth,height=0.9\textwidth]{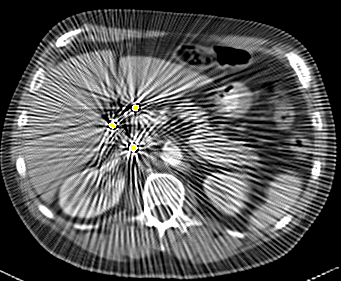}
        \end{subfigure} &
        \begin{subfigure}[b]{.18\linewidth}
            \setlength{\abovecaptionskip}{2pt}
            \includegraphics[width=\textwidth,height=0.9\textwidth]{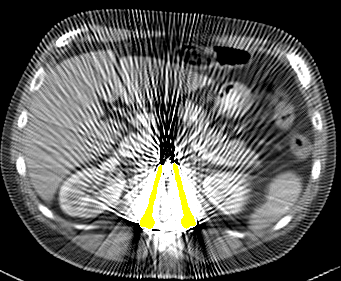}
        \end{subfigure} \\
        \end{tabular}
    \end{tabular}
    \caption{Sample simulated metal artifact on patient CT. The X-ray spectrum is shown in the upper-left corner. Metallic implants are colored in yellow for better visualization.}
    \label{fig:simulation}
\end{figure*}
\textbf{Radon Consistency Loss.} With the  differentiable RIL, we introduce the following Radon consistency (RC) loss to penalize secondary artifacts in $\hat{X} = f_R(Y_{out})$ after reconstruction.
\begin{align}
    \mathcal{L}_{RC} &= \lVert f_R(Y_{out}) - X_{gt} \rVert_1,
\end{align}
where $X_{gt}$ is the ground truth CT image without metal artifact.

\textbf{Difference from DL-based Reconstruction.} Our RIL is designed to combine the image formation process (CT reconstruction) with deep neural networks and achieve improved MAR by dual-domain consistency learning. Methods in~\cite{Wurfl-16-MICCAI,Jin-17-TIP,Adler-18-TMI,Zhang-18-TMI} target \emph{image formation via deep learning}, which is not the main focus of this work.

\subsection{Image Enhancement Network}
Since our ultimate goal is to reduce visually undesirable artifacts in image domain, we further apply a U-Net $\mathcal{G}_i$ to enhance $\hat{X}$ by residual learning:
\begin{align}
    X_{out} &= X_{LI} + \mathcal{G}_i(\hat{X}, X_{LI}),
\end{align}
where $X_{LI}=f_R(Y_{LI})$ is reconstructed from $Y_{LI}$, the linearly interpolated sinogram.
$\mathcal{G}_i$ is also optimized by $L_1$ loss.
\begin{align}
    \mathcal{L}_{\mathcal{G}_i} &= \lVert X_{out} - X_{gt} \rVert_1.
\end{align}
The full objective function of our model is:
\begin{align}
    \mathcal{L} &= \mathcal{L}_{\mathcal{G}_s}+\mathcal{L}_{RC}+\mathcal{L}_{\mathcal{G}_i}.
    \label{eq:loss}
\end{align}
One could tune and balance each term in~\eqref{eq:loss} for better performance. However, we found that the default setting works sufficiently well.
\section{Experimental Results}

Following the de facto practice in the literature~\cite{Zhang-18-MAR-TMI}, our evaluations consider simulated metal artifacts on real patient CTs. Various effects are considered including polychromatic X-ray, partial volume effect, and Poisson noise. The simulated artifacts exhibit complicated structures and cannot be easily modelled by a very deep CNN. All the compared approaches are evaluated on the same dataset, and superior performance is achieved by our method. Evaluations on clinical data is presented in the supplementary material.


\textbf{Metal Artifact Dataset.}
Recently, Yan et al.~\cite{Yan-18-lesion} released a large-scale CT dataset DeepLesion for lesion detection. Due to its high diversity and quality, we use a subset of images from DeepLesion to synthesize metal artifact. 4,000 images from 320 patients are used in the training set and 200 images from 12 patients are used in the test set. All images are resized to $416 \times 416$. We collect a total of 100 metal shapes. 90 metal shapes are paired with the 4,000 images, yielding 360,000 combinations in the training set. 10 metal shapes are paired with the 200 images, yielding 2,000 combinations in the test set. In the training set, the sizes of the metal implants range from 16 to 4967 pixels. In the test set, the sizes of the metal implants range from 32 to 2054 pixels.

We adopt similar procedures as in~\cite{Zhang-18-MAR-TMI} to synthesize metal-corrupted sinograms and CT images. We assume a polychromatic X-ray source with spectrum $\eta(E)$ in Figure~\ref{fig:simulation}. To simulate Poisson noise in the sinogram, we assume the incident X-ray has $2 \times 10^7$ photons. Metal partial volume effect is also considered. The distance from the X-ray source to the rotation center is set to 39.7cm, and 320 projection views are uniformly spaced between 0-360 degrees. The resulting sinograms have size $321 \times 320$. Figure~\ref{fig:simulation} shows some sample images with simulated metal artifacts.

\textbf{Evaluation Metrics.} We choose peak signal-to-noise ratio (PSNR) and structured similarity index (SSIM) for quantitative evaluations. In DeepLesion, each CT image is provided with a dynamic range, within which the tissues are clearly discernible. We use the dynamic range as the peak signal strength when calculating PSNR.

\begin{table*}[t!]
\small
\centering 
\begin{tabular}{l | ccccc | c }
\toprule
PSNR(dB)/SSIM & \multicolumn{5}{|c|}{Large Metal $\xrightarrow{\hspace*{2.4cm}}$ Small Metal} & Average \\
\midrule
A) $\mbox{SE-Net}_0$  & $22.88/0.7850$ & $24.52/0.8159$ & $27.38/0.8438$ & $28.61/0.8549$ & $28.93/0.8581$ & $26.46/0.8315$  \\
B) SE-Net & $23.06/0.7868$ & $24.71/0.8178$ & $27.66/0.8463$ & $28.91/0.8575$ & $29.19/0.8604$ & $26.71/0.8337$  \\
C) IE-Net & $27.54/0.8840$ & $29.49/0.9153$ & $31.96/0.9368$ & $34.38/0.9498$ & $33.90/0.9489$ & $31.45/0.9269$  \\
D) $\mbox{SE-Net}_0+$IE-Net & $28.46/0.8938$ & $30.67/0.9232$ & $33.71/0.9458$ & $36.17/0.9576$ & $35.74/0.9571$ & $32.95/0.9355$  \\
E) SE-Net$+$IE-Net & $28.28/0.8921$ & $30.49/0.9221$ & $33.76/0.9456$ & $36.26/0.9576$ & $36.01/0.9574$ & $32.96/0.9350$  \\
F) $\mbox{SE-Net}_0+$IE-Net$+$RCL & $28.97/0.8970$ & $31.14/0.9254$ & $34.21/0.9476$ & $36.58/0.9590$ & $36.15/0.9586$ & $33.41/0.9375$  \\
G) SE-Net$+$IE-Net$+$RCL & ${29.02}/{0.8972}$ & ${31.12}/{0.9256}$ &  ${34.32}/{0.9481}$ & ${36.72}/{0.9595}$ & ${36.36}/{0.9592}$ & ${33.51}/{0.9379}$  \\
\bottomrule
\end{tabular}
\vspace{-2pt}
\caption{Quantitative evaluations for different components in DuDoNet.}
\label{exp:ablation_table}
\end{table*}
\textbf{Implementation Details.}
We implement our model using the PyTorch~\cite{paszke2017automatic} framework. All the sinograms have size $321 \times 320$, and all the CT images have size $416 \times 416$. To train the model, we use the Adam~\cite{kingma2014adam} optimizer with $(\beta_1, \beta_2) = (0.5, 0.999)$, and a batch size of 8. The learning rate starts from $2 \times 10^{-4}$, and is halved for every 30 epochs. The model is trained on two Nvidia 1080Ti for 380 epochs.

\subsection{Ablation Study}
In this section, we evaluate the effectiveness of different components in the proposed approach. Performance is evaluated on the artifact-reduced CT images. When evaluating SE-Nets without image domain refinement, we use the reconstructed CT images $\hat{X}$. We experiment on the following configurations:
\begin{enumerate}[label=\Alph*),itemsep=1 pt]
    \item $\mbox{SE-Net}_0$: The sinogram enhancement network without mask pyramid network.
    \item SE-Net: The full sinogram enhancement module.
    \item IE-Net: Image enhancement module. IE-Net is applied to enhance $X_{LI}$ without $\hat{X}$.
    \item $\mbox{SE-Net}_0+$IE-Net: Dual domain learning with $\mbox{SE-Net}_0$ and IE-Net.
    \item SE-Net$+$IE-Net: Dual domain learning with SE-Net and IE-Net.
    \item $\mbox{SE-Net}_0+$IE-Net$+$RCL: Dual domain learning with Radon consistency loss.
    \item SE-Net$+$IE-Net$+$RCL: Our full network. 
\end{enumerate}
Notice that the configurations including $\mbox{SE-Net}_0$, SE-Net and IE-Net are single domain enhancement approaches. 

Table~\ref{exp:ablation_table} summarizes the performance of different models. Since there are totally 10 metal implants in the test set, for conciseness, we group the results according to the size of metal implants. The sizes of the 10 metal implants are: [2054, 879, 878, 448, 242, 115, 115, 111, 53, 32] in pixels. We simply put every two masks into one group.

\begin{figure}[!h]
    \vspace{-0.3cm}
    \setlength{\abovecaptionskip}{2pt}
    \centering
    \begin{subfigure}[b]{0.32\linewidth}
        \setlength{\abovecaptionskip}{2pt}
        \includegraphics[width=\textwidth,height=0.82\textwidth]{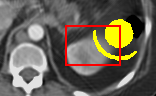}
        \caption*{Without RC loss}
    \end{subfigure}
    \begin{subfigure}[b]{0.32\linewidth}
        \setlength{\abovecaptionskip}{2pt}
        \includegraphics[width=\textwidth,height=0.82\textwidth]{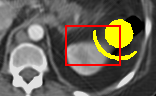}
        \caption*{With RC loss}
    \end{subfigure}
    \begin{subfigure}[b]{0.32\linewidth}
        \setlength{\abovecaptionskip}{2pt}
        \includegraphics[width=\textwidth,height=0.82\textwidth]{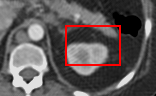}
        \caption*{Ground Truth}
    \end{subfigure}
    \caption{Visual comparisons between models without RC loss (E in Table~\ref{exp:ablation_table}) and our full model (G in Table~\ref{exp:ablation_table}).}
    \label{fig:cons}
\end{figure}

From E and G, it is clear that the use of the RC loss improves the performance over all metal sizes for at least 0.3 dB. In Figure~\ref{fig:cons}, the model trained with RC loss better recovers the shape of the organ.

\begin{figure}[!h]
    \vspace{-0.1cm}
    \setlength{\abovecaptionskip}{2pt}
    \centering
    \begin{subfigure}[b]{0.32\linewidth}
        \setlength{\abovecaptionskip}{2pt}
        \includegraphics[width=\textwidth,height=0.82\textwidth]{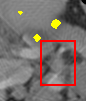}
        \caption*{Without MP}
    \end{subfigure}
    \begin{subfigure}[b]{0.32\linewidth}
        \setlength{\abovecaptionskip}{2pt}
        \includegraphics[width=\textwidth,height=0.82\textwidth]{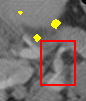}
        \caption*{With MP}
    \end{subfigure}
    \begin{subfigure}[b]{0.32\linewidth}
        \setlength{\abovecaptionskip}{2pt}
        \includegraphics[width=\textwidth,height=0.82\textwidth]{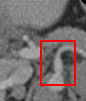}
        \caption*{Ground Truth}
    \end{subfigure}
    \caption{Visual comparisons between models without MP (F in Table~\ref{exp:ablation_table}) and our full model (G in Table~\ref{exp:ablation_table}).}
    \label{fig:mask}
    \vspace{-0.1cm}
\end{figure}

From F and G, we observe an interesting trend that the proposed mask pyramid architecture results in $\sim$0.2 dB gain when the metal size is small, and the performance is nearly identical when the metal is large. The reason is that the mask pyramid retains metal information across multiple scales. Figure~\ref{fig:mask} demonstrates that in the proximity of small metal implants, the model with mask pyramid recovers the fine details.

\begin{figure}[!h]
    \setlength{\abovecaptionskip}{2pt}
    \centering
    \begin{subfigure}[b]{0.32\linewidth}
        \setlength{\abovecaptionskip}{2pt}
        \includegraphics[width=\textwidth,height=0.82\textwidth]{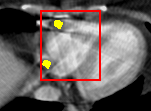}
        \caption*{$X_{LI}$}
    \end{subfigure}
    \begin{subfigure}[b]{0.32\linewidth}
        \setlength{\abovecaptionskip}{2pt}
        \includegraphics[width=\textwidth,height=0.82\textwidth]{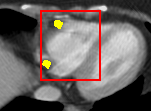}
        \caption*{IE-Net}
    \end{subfigure}
    \begin{subfigure}[b]{0.32\linewidth}
        \setlength{\abovecaptionskip}{2pt}
        \includegraphics[width=\textwidth,height=0.82\textwidth]{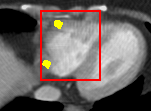}
        \caption*{IE-Net-RDN}
    \end{subfigure}
     \\
    \vspace{2pt}
    \begin{subfigure}[b]{0.32\linewidth}
        \setlength{\abovecaptionskip}{2pt}
        \includegraphics[width=\textwidth,height=0.82\textwidth]{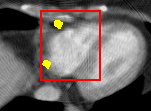}
        \caption*{$\hat{X}$}
    \end{subfigure}
    \begin{subfigure}[b]{0.32\linewidth}
        \setlength{\abovecaptionskip}{2pt}
        \includegraphics[width=\textwidth,height=0.82\textwidth]{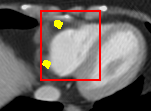}
        \caption*{$X_{out}$}
    \end{subfigure}
    \begin{subfigure}[b]{0.32\linewidth}
        \setlength{\abovecaptionskip}{2pt}
        \includegraphics[width=\textwidth,height=0.82\textwidth]{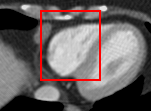}
        \caption*{Ground Truth}
    \end{subfigure}
    \caption{Visual comparisons between models without SE-Net (top row IE-Net and IE-Net-RDN) and our full model (bottom row $\hat{X}$ and $X_{out}$).}
    \label{fig:SEM}
    \vspace{-0.2cm}
\end{figure}

\begin{table*}[t!]
\small
\centering 
\begin{tabular}{l | ccccc | c }
\toprule
 PSNR(dB)/SSIM& \multicolumn{5}{|c|}{Large Metal $\xrightarrow{\hspace*{3.4cm}}$ Small Metal} & Average \\
\midrule
LI  & $20.20/0.8236$ & $22.35/0.8686$ & $26.76/0.9098$ & $28.50/0.9252$ & $29.53/0.9312$ & $25.47/0.8917$  \\
NMAR & $21.95/0.8333$ & $24.43/0.8813$ & $28.63/0.9174$ & $30.84/0.9281$ & $31.69/0.9402$ & $27.51/0.9001$  \\
cGAN-CT & $26.71/0.8265$ & $24.71/0.8507$ & $29.80/0.8911$ & $31.47/0.9104$ & $27.65/0.8876$ & $28.07/0.8733$  \\
RDN-CT & $\underline{28.61}/0.8668$ & $\underline{28.78}/0.9027$ & $\underline{32.40}/0.9264$ & $\underline{34.95}/0.9446$ & $\underline{34.00}/0.9376$ & $\underline{31.74}/0.9156$  \\
CNNMAR & $23.82/\underline{0.8690}$ & $26.78/\underline{0.9097}$ & $30.92/\underline{0.9394}$ & $32.97/\underline{0.9513}$ & $33.11/\underline{0.9520}$ & $29.52/\underline{0.9243}$  \\
DuDoNet (Ours) & $\bm{29.02}/\bm{0.8972}$ & $\bm{31.12}/\bm{0.9256}$ & $\bm{34.32}/\bm{0.9481}$ & $\bm{36.72}/\bm{0.9595}$ & $\bm{36.36}/\bm{0.9592}$ & $\bm{33.51}/\bm{0.9379}$  \\
\bottomrule
\end{tabular}
\vspace{-1pt}
\caption{Quantitative evaluation of MAR approaches in terms of PSNR and SSIM.}
\label{exp:sota_table}
\vspace{-1pt}
\end{table*}
\begin{figure*}[!t]
    \setlength{\abovecaptionskip}{3pt}
    \setlength{\tabcolsep}{1pt}
    \begin{tabular}[b]{cc}
        \begin{subfigure}[b]{0.28\linewidth}
            \includegraphics[width=\linewidth,height=0.84\textwidth]{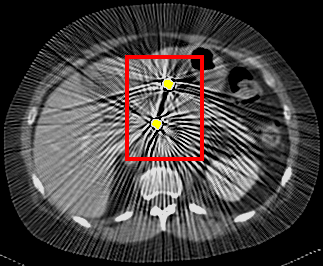}
        \caption{Small metallic implants.}
        \label{fig:small_ma}
        \end{subfigure}
        &
        \begin{tabular}[b]{c c c c}
            \begin{subfigure}[b]{0.18\linewidth}
                \setlength{\abovecaptionskip}{3pt}
                \includegraphics[width=\linewidth,height=0.5\textwidth,cframe=red]{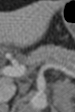}
                \caption*{\stackanchor{Ground Truth}{PSNR/SSIM}}
            \end{subfigure}
            &
            \begin{subfigure}[b]{0.18\linewidth}
                \setlength{\abovecaptionskip}{3pt}
                \includegraphics[width=\linewidth,height=0.5\textwidth,cframe=red]{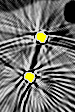}
                \caption*{\stackanchor{With Metal Artifact}{10.98/0.1485}}
            \end{subfigure}
            &
            \begin{subfigure}[b]{0.18\linewidth}
                \setlength{\abovecaptionskip}{2pt}
                \includegraphics[width=\linewidth,height=0.5\textwidth,cframe=red]{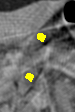}
                \caption*{\stackanchor{LI}{20.62/0.5462}}
            \end{subfigure}
            &
            \begin{subfigure}[b]{0.18\linewidth}
                \setlength{\abovecaptionskip}{2pt}
                \includegraphics[width=\linewidth,height=0.5\textwidth,cframe=red]{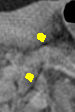}
                \caption*{\stackanchor{NMAR}{23.21/0.6336}}
            \end{subfigure} \\ 
            \begin{subfigure}[b]{0.18\linewidth}
                \setlength{\abovecaptionskip}{2pt}
                \includegraphics[width=\linewidth,height=0.5\textwidth,cframe=red]{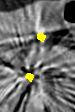}
                \caption*{\stackanchor{cGAN-CT}{15.12/0.2678}}
            \end{subfigure}
            &
            \begin{subfigure}[b]{0.18\linewidth}
                \setlength{\abovecaptionskip}{2pt}
                \includegraphics[width=\linewidth,height=0.5\textwidth,cframe=red]{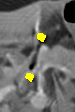}
                \caption*{\stackanchor{RDN-CT}{20.88/0.5353}}
            \end{subfigure}
            &
            \begin{subfigure}[b]{0.18\linewidth}
                \setlength{\abovecaptionskip}{2pt}
                \includegraphics[width=\linewidth,height=0.5\textwidth,cframe=red]{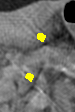}
                \caption*{\stackanchor{CNNMAR}{23.11/0.6405}}
            \end{subfigure}
            &
            \begin{subfigure}[b]{0.18\linewidth}
                \setlength{\abovecaptionskip}{3pt}
                \includegraphics[width=\linewidth,height=0.5\textwidth,cframe=red]{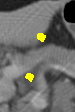}
                \caption*{\stackanchor{DuDoNet}{\textbf{26.91}/\textbf{0.7258}}}
            \end{subfigure}
        \end{tabular} \\ 
        \begin{subfigure}[b]{0.28\linewidth}
            \includegraphics[width=\linewidth,height=0.84\textwidth]{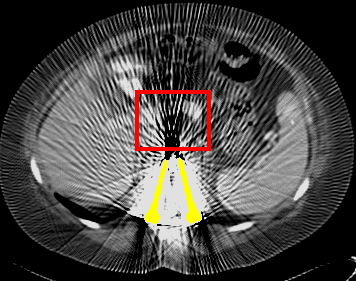}
        \caption{Medium metallic implants.}
        \label{fig:medium_ma}
        \end{subfigure}
        & 
        \begin{tabular}[b]{c c c c}
            \begin{subfigure}[b]{0.18\linewidth}
                \setlength{\abovecaptionskip}{3pt}
                \includegraphics[width=\linewidth,height=0.5\textwidth,cframe=red]{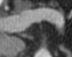}
                \caption*{\stackanchor{Ground Truth}{PSNR/SSIM}}
            \end{subfigure}
            &
            \begin{subfigure}[b]{0.18\linewidth}
                \setlength{\abovecaptionskip}{3pt}
                \includegraphics[width=\linewidth,height=0.5\textwidth,cframe=red]{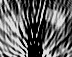}
                \caption*{\stackanchor{With Metal Artifact}{9.67/0.1137}}
            \end{subfigure}
            &
            \begin{subfigure}[b]{0.18\linewidth}
                \setlength{\abovecaptionskip}{2pt}
                \includegraphics[width=\linewidth,height=0.5\textwidth,cframe=red]{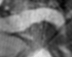}
                \caption*{\stackanchor{LI}{18.36/0.6628}}
            \end{subfigure}
            &
            \begin{subfigure}[b]{0.18\linewidth}
                \setlength{\abovecaptionskip}{2pt}
                \includegraphics[width=\linewidth,height=0.5\textwidth,cframe=red]{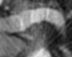}
                \caption*{\stackanchor{NMAR}{19.08/0.6697}}
            \end{subfigure} \\ 
            \begin{subfigure}[b]{0.18\linewidth}
                \setlength{\abovecaptionskip}{2pt}
                \includegraphics[width=\linewidth,height=0.5\textwidth,cframe=red]{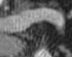}
                \caption*{\stackanchor{cGAN-CT}{28.15/0.7328}}
            \end{subfigure}
            &
            \begin{subfigure}[b]{0.18\linewidth}
                \setlength{\abovecaptionskip}{2pt}
                \includegraphics[width=\linewidth,height=0.5\textwidth,cframe=red]{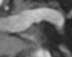}
                \caption*{\stackanchor{RDN-CT}{21.52/0.6966}}
            \end{subfigure}
            &
            \begin{subfigure}[b]{0.18\linewidth}
                \setlength{\abovecaptionskip}{2pt}
                \includegraphics[width=\linewidth,height=0.5\textwidth,cframe=red]{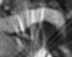}
                \caption*{\stackanchor{CNNMAR}{19.66/0.6370}}
            \end{subfigure}
            &
            \begin{subfigure}[b]{0.18\linewidth}
                \setlength{\abovecaptionskip}{3pt}
                \includegraphics[width=\linewidth,height=0.5\textwidth,cframe=red]{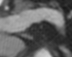}
                \caption*{\stackanchor{DuDoNet}{\textbf{28.72}/\textbf{0.8108}}}
            \end{subfigure}
        \end{tabular} \\ 
        \begin{subfigure}[b]{0.28\linewidth}
            \includegraphics[width=\linewidth,height=0.84\textwidth]{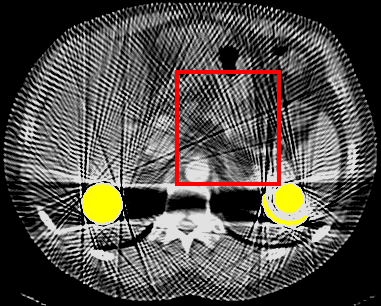}
        \caption{Large metallic implants.}
        \label{fig:large_ma}
        \end{subfigure}
        & 
        \begin{tabular}[b]{c c c c}
            \begin{subfigure}[b]{0.18\linewidth}
                \setlength{\abovecaptionskip}{3pt}
                \includegraphics[width=\linewidth,height=0.5\textwidth,cframe=red]{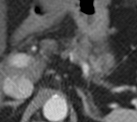}
                \caption*{\stackanchor{Ground Truth}{PSNR/SSIM}}
            \end{subfigure}
            &
            \begin{subfigure}[b]{0.18\linewidth}
                \setlength{\abovecaptionskip}{3pt}
                \includegraphics[width=\linewidth,height=0.5\textwidth,cframe=red]{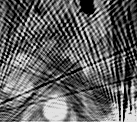}
                \caption*{\stackanchor{With Metal Artifact}{12.15/0.1519}}
            \end{subfigure}
            &
            \begin{subfigure}[b]{0.18\linewidth}
                \setlength{\abovecaptionskip}{2pt}
                \includegraphics[width=\linewidth,height=0.5\textwidth,cframe=red]{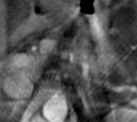}
                \caption*{\stackanchor{LI}{19.27/0.6260}}
            \end{subfigure}
            &
            \begin{subfigure}[b]{0.18\linewidth}
                \setlength{\abovecaptionskip}{2pt}
                \includegraphics[width=\linewidth,height=0.5\textwidth,cframe=red]{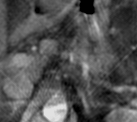}
                \caption*{\stackanchor{NMAR}{20.20/0.6597}}
            \end{subfigure} \\ 
            \begin{subfigure}[b]{0.18\linewidth}
                \setlength{\abovecaptionskip}{2pt}
                \includegraphics[width=\linewidth,height=0.5\textwidth,cframe=red]{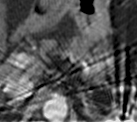}
                \caption*{\stackanchor{cGAN-CT}{18.68/0.4460}}
            \end{subfigure}
            &
            \begin{subfigure}[b]{0.18\linewidth}
                \setlength{\abovecaptionskip}{2pt}
                \includegraphics[width=\linewidth,height=0.5\textwidth,cframe=red]{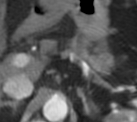}
                \caption*{\stackanchor{RDN-CT}{26.28/0.6946}}
            \end{subfigure}
            &
            \begin{subfigure}[b]{0.18\linewidth}
                \setlength{\abovecaptionskip}{2pt}
                \includegraphics[width=\linewidth,height=0.5\textwidth,cframe=red]{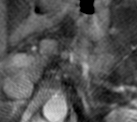}
                \caption*{\stackanchor{CNNMAR}{20.92/0.6916}}
            \end{subfigure}
            &
            \begin{subfigure}[b]{0.18\linewidth}
                \setlength{\abovecaptionskip}{3pt}
                \includegraphics[width=\linewidth,height=0.5\textwidth,cframe=red]{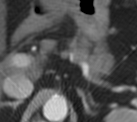}
                \caption*{\stackanchor{DuDoNet}{\textbf{27.31}/\textbf{0.7947}}}
            \end{subfigure}
        \end{tabular}
    \end{tabular}
    \caption{Visual comparisons on MAR for different types of metallic implants.} \label{fig:sota}
\end{figure*}

\textbf{Effect of Dual Domain Learning.} In the proposed framework, IE-Net enhances $X_{LI}$ by fusing information from SE-Net. We study the effect of dual domain learning by visually comparing our full pipeline (G in Table~\ref{exp:ablation_table}) with single domain enhancement IE-Net (C in Table~\ref{exp:ablation_table}). In addition to the U-Net architecture, we also consider IE-Net with RDN architecture, which is denoted as IE-Net-RDN. Visual comparisons are shown in Figure~\ref{fig:SEM}. We observe that single domain models IE-Net and IE-Net-RDN fail to recover corrupted organ boundaries in $X_{LI}$. In our dual domain refinement network, SE-Net first recovers inconsistent sinograms and reduces secondary artifacts as in $\hat{X}$. IE-Net then refines $\hat{X}$ to recover the fine details.

\textbf{Effect of LI sinogram.} The inputs to our network are the linear interpolated sinogram $Y_{LI}$ and its reconstructed CT $X_{LI}$. One possible alternative is to directly input the metal corrupted sinogram and CT, and let the network learn to restore the intense artifacts. However, we experimentally found out this alternative approach does not perform well. Metal shadows and streaking artifacts are not fully suppressed.

\subsection{Comparison with State-of-the-Art Methods}
In this section, we compare our model with the following methods: LI~\cite{Kalender-87-LI}, NMAR~\cite{Meyer-10-nmar}, cGAN-CT~\cite{Wang-18-MAR-MICCAI}, RDN-CT~\cite{zhang-18-RDN} and CNNMAR~\cite{Zhang-18-MAR-TMI}. We use cGAN-CT to refer the approach by Wang et al.~\cite{Wang-18-MAR-MICCAI} which applies cGAN for image domain MAR. RDN~\cite{zhang-18-RDN} was originally proposed for image super-resolution (SR). The fundamental building unit of RDN is the residual dense block (RDB). Recently, it has been shown that by stacking multiple RDBs or its variant, the residual in residual dense blocks (RRDBs)~\cite{wang-18-esrgan}, local details in natural images can be effectively recovered. We build a very deep architecture with 10 RDBs ($\sim$80 conv layers) for direct image domain enhancement, which is denoted by RDN-CT. Specifically, we select $D=10, C=8, G=64$, following the notations in~\cite{zhang-18-RDN}. Inputs to RDN-CT are $128 \times 128$ patches.

\textbf{Quantitative Comparisons.} Table~\ref{exp:sota_table} shows quantitative comparisons. We observe that the state-of-the-art sinogram inpainting approach CNNMAR achieves higher SSIM than image enhancement approaches (e.g. RDN and cGAN-CT) especially when the size of metal is small. The reason is that sinogram inpainting only modifies data within the metal trace and recovers the statistics reasonably well. In most of the cases, CNNMAR also outperforms cGAN-CT in terms of PSNR. However, when CNN is sufficiently deep (e.g. RDN-CT), image enhancement approaches generally achieve higher PSNR. Our dual domain learning approach jointly restores sinograms and CT images, which attains the best performance in terms of both PSNR and SSIM {\it consistently in all categories}.

\textbf{Visual Comparisons.} Figure~\ref{fig:sota} shows visual comparisons. Figure~\ref{fig:small_ma} considers metal artifacts resulted from two small metallic implants. From the zoomed figure (with metal artifact), we can perceive severe streaking artifacts and intense metal shadows between the two implants. We observe that sinogram inpainting approaches such as LI, NMAR and CNNMAR effectively reduce metal shadows. However, fine details are either corrupted by secondary artifacts as in LI or blurred as in NMAR and CNNMAR. Image domain approaches such as cGAN-CT and RDN-CT produce sharper CT images but fail to suppress metal shadows. Our method effectively reduces metal shadows and at the same time retains fine details. Figure~\ref{fig:medium_ma} shows a degraded CT image with long metal implants. We observe similar trend that sinogram inpainting approaches do not perform well in regions with intense streaking artifact. In this example, image domain methods reduce most of the artifacts. It is possibly due to that fact that the pattern of the artifact in Figure~\ref{fig:medium_ma} is monotonous compared to Figures~\ref{fig:small_ma} and~\ref{fig:large_ma}. However, noticeable speckle noise is present in the result by cGAN-CT, and RDN-CT does not fully recover details in the middle.  Figure~\ref{fig:large_ma} considers metal artifacts result from two large metallic implants. Likewise, sinogram inpainting methods and direct image domain enhancement have limited capability of suppressing metal artifacts. More visual comparisons are presented in the supplemental material. 

\subsection{Running Time Comparisons}
On an Nvidia 1080Ti GPU, it takes 0.24 ms for RIL to reconstruct a sinogram of size $321 \times 320$ to a CT image of size $416 \times 416$, and 11.40 ms for back-propagation of gradients. RIL requires 16 MB of memory for forward pass and 25 MB for back-propagation. In Table~\ref{exp:running_time} we compare the running time of different MAR approaches. With the running time of LI included, DuDoNet runs almost $4\times$ faster than the very deep architecture RDN while achieving superior performance.
\begin{table}[h!]
\small
\centering
\setlength{\tabcolsep}{3pt}
\begin{tabular}{cccccc}
\toprule
\stackanchor{LI}{~\cite{Kalender-87-LI}} & \stackanchor{NMAR}{~\cite{Meyer-10-nmar}} & \stackanchor{cGAN-CT}{\cite{Wang-18-MAR-MICCAI}} & \stackanchor{RDN-CT}{\cite{zhang-18-RDN}}  & \stackanchor{CNNMAR}{\cite{Zhang-18-MAR-TMI}} & \stackanchor{DuDoNet}{(Ours)}  \\
\midrule
0.0832 &  0.4180 & 0.0365 &  0.5150 & 0.6043  &  0.1335  \\
\bottomrule
\end{tabular}
\caption{Comparison of running time measured in seconds.}
\label{exp:running_time}
\end{table}

\section{Conclusion}
In this paper, we present the Dual Domain Network for metal artifact reduction. In particular, we propose to jointly improve sinogram consistency and refine CT images through a novel Radon inversion layer and a Radon consistency loss, along with a mask pyramid U-Net. Experimental evaluations demonstrate that while state-of-the-art MAR methods suffer from secondary artifacts and very-deep neural networks have limited capability of directly reducing metal artifacts in image domain, our dual-domain model can effectively suppress metal shadows and recover details for CT images. At the same time, our network is computationally more efficient. Future work includes investigating the potential of the dual-domain learning framework for other signal recovery tasks, such as super-resolution, noise reduction, and CT reconstruction from sparse X-ray projections.

{\vspace{0.5em} \noindent \bf Acknowledgements:} This research was supported by MURI from the Army Research Office under the Grant No. W911NF-17-1-0304, NSF award \#17228477, and the Morris K. Udall Center of Excellence in Parkinson's Disease Research by NIH.
\appendix
\section{Fanbeam CT Geometry}
\begin{figure}[!h]
    \centering
      \includegraphics[width=0.7\linewidth]{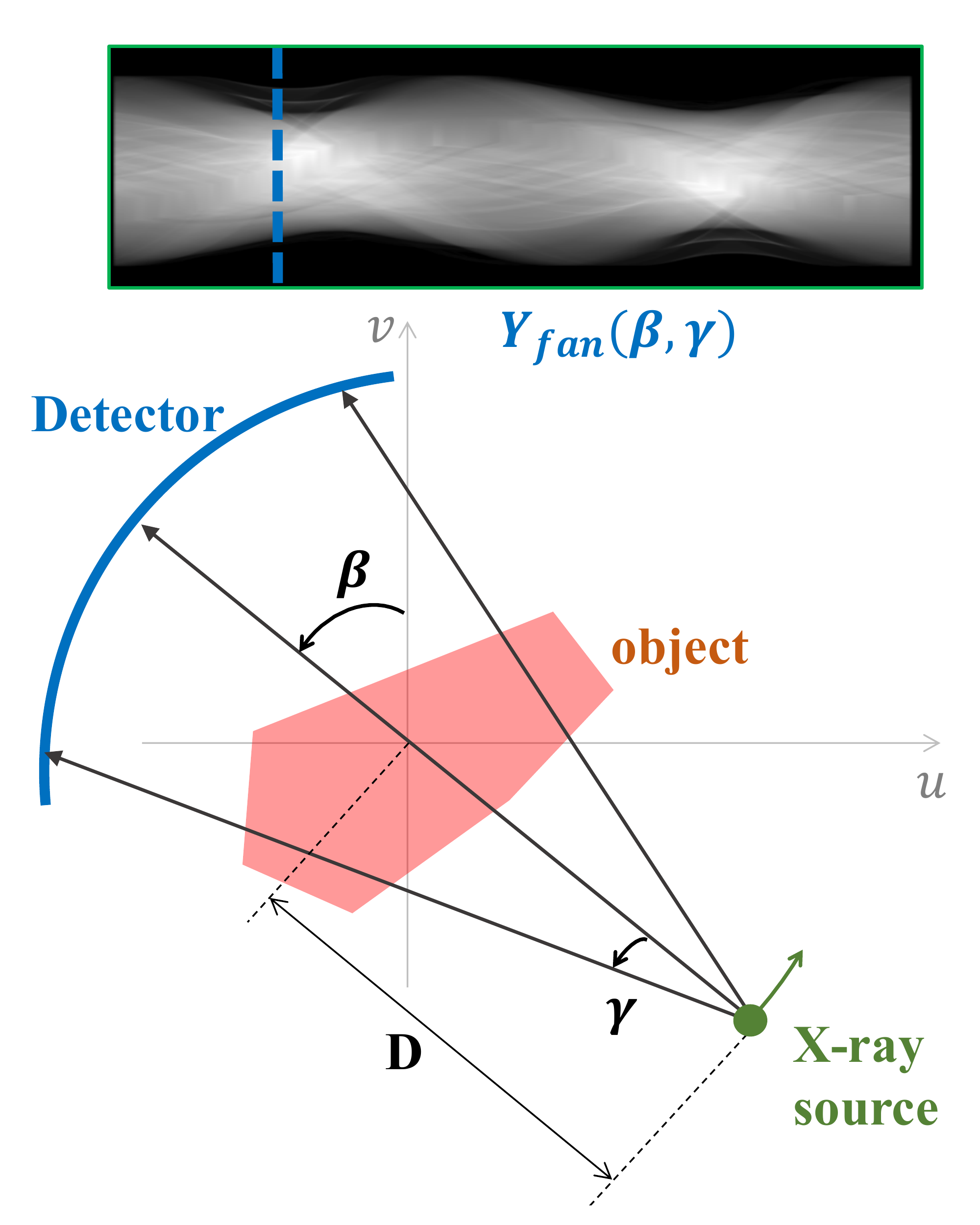}
    \caption{Fanbeam CT geometry.}
    \label{fig:fanbeam}
\end{figure}

Figure~\ref{fig:fanbeam} illustrates the general fanbeam CT geometry. The X-ray source and the arc detector rotate with respect to the origin. The distance between the X-ray source and the origin is $D$. For each projection angle $\beta$, the arc detector receives the X-rays transmitted from the object. The intensity values received by the detector is represented as a 1D signal with independent variable $\gamma$. As shown in the top of Figure~\ref{fig:fanbeam}, the sinogram data $Y_{fan}(\beta, \gamma)$ consists of the 1D signals received in different projection angles $\beta$.

\section{Implementation Details}

\subsection{Network Architecture}
The proposed DuDoNet consists of SE-Net and IE-Net. The architecture of SE-Net is presented in Table~\ref{table:se-net}. The architecture of IE-Net is identical to U-Net~\cite{ronneberger-15-unet}. $N_c$ denotes the number of output channels. $\mathcal{M}_t \downarrow k$ represents the sinogram mask down-sized to $1/k$. All downsampling convolution layers except for the first layer use leaky ReLU activation function with $\alpha=0.2$. All upsampling convolution layers except for the last layer use ReLU activation function. We use `K\#-C\#-S\#-P\#' to denote the configuration of the convolution layers, where `K', `C', `S' and `P' stand for the kernel, channel, stride and padding size, respectively.
\begin{table}[h!]
\centering 
\begin{tabular}{l | c | l}
Name & $N_{c}$ & Description \\
\hline
INPUT & $2$ & Input sinogram and $\mathcal{M}_t$\\ 
DOWN\_CONV0 & $64$ & K4-C64-S2-P1 \\
CONCAT0 & $65$ & Concatenate $\mathcal{M}_t \downarrow 2$ \\
DOWN\_CONV1 & $128$ & K4-C128-S2-P1 \\
CONCAT1 & $129$ & Concatenate $\mathcal{M}_t \downarrow 4$ \\
DOWN\_CONV2 & $256$ & K4-C256-S2-P1 \\
CONCAT2 & $257$ & Concatenate $\mathcal{M}_t \downarrow 8$ \\
DOWN\_CONV3 & $512$ & K4-C512-S2-P1 \\
CONCAT3 & $513$ & Concatenate $\mathcal{M}_t \downarrow 16$ \\
DOWN\_CONV4 & $512$ & K4-C512-S2-P1 \\
CONCAT4 & $513$ & Concatenate $\mathcal{M}_t \downarrow 32$ \\
\midrule
UPSAMPLE5 & $513$ &  \\
UP\_CONV5 & $512$ & K3-C512-S1-P1 \\
CONCAT5 & $(512+513)$ & Concatenate CONCAT3 \\
UPSAMPLE6 & $(512+513)$ &  \\
UP\_CONV6 & $256$ & K3-C256-S1-P1 \\
CONCAT6 & $(256+257)$ & Concatenate CONCAT2 \\
UPSAMPLE7 & $(256+257)$ &  \\
UP\_CONV7 & $128$ & K3-C128-S1-P1 \\
CONCAT7 & $(128+129)$ & Concatenate CONCAT1 \\
UPSAMPLE8 & $(128+129)$ &  \\
UP\_CONV8 & $64$ & K3-C64-S1-P1 \\
CONCAT8 & $(64+65)$ & Concatenate CONCAT0 \\
UPSAMPLE9 & $(64+65)$ &  \\
UP\_CONV9 & $1$ & K3-C1-S1-P1 \\
\end{tabular}
\caption{Network architecture of SE-Net.}
\label{table:se-net}
\end{table}

\subsection{Radon Inversion Layer}
We implement our Radon Inversion Layer (RIL) in PyTorch~\cite{paszke2017automatic} and CUDA. In the following, we detail our implementation.

\begin{figure}[!h]
    \centering
      \includegraphics[width=0.7\linewidth]{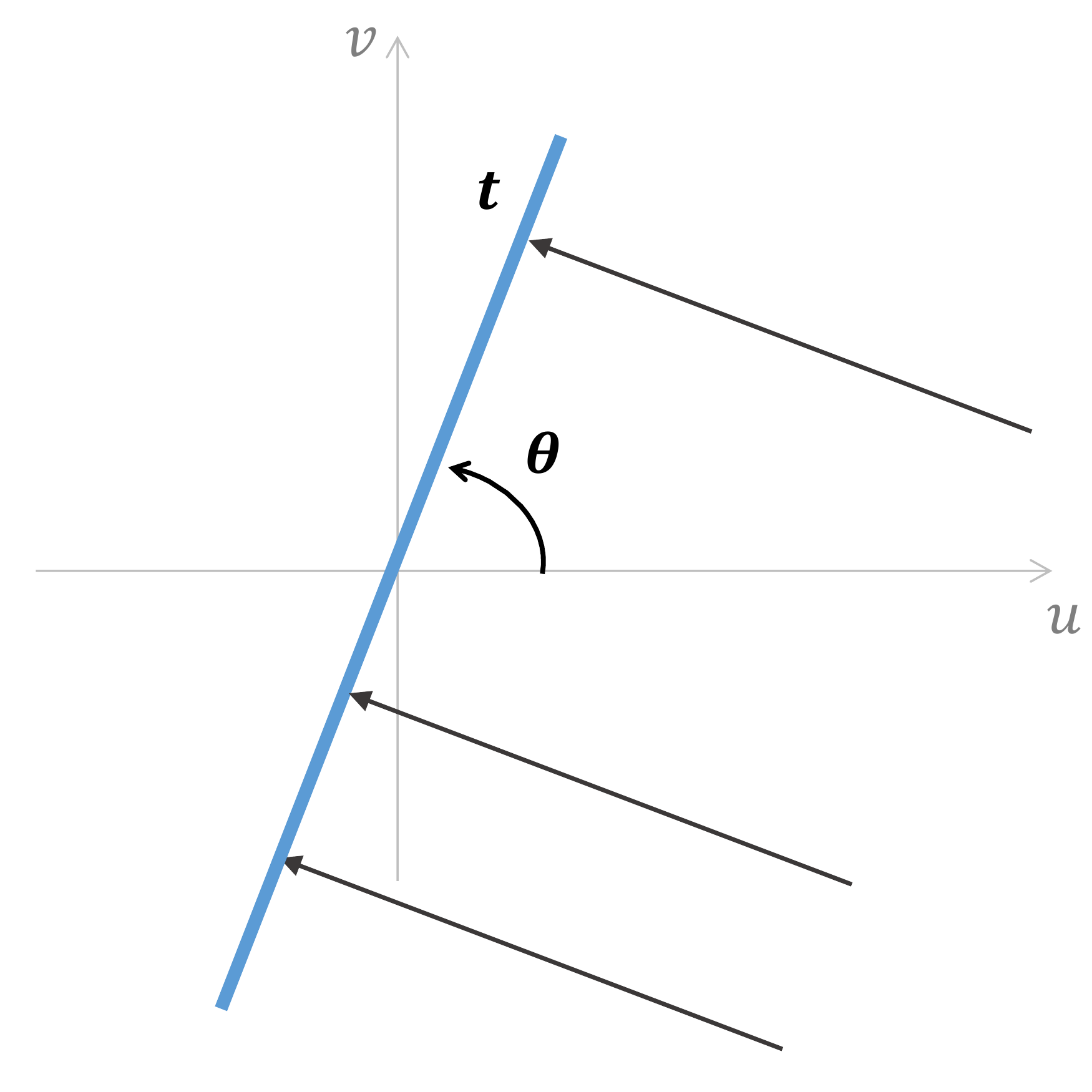}
    \caption{Parallel-beam CT geometry.}
    \label{fig:para}
\end{figure}

RIL consists of three modules: (1) parallel-beam conversion module, (2) Ram-Lak filtering module and (3) backprojection module. Given a fanbeam sinogram $Y_{fan}(\beta, \gamma)$, we first convert it to a parallel beam sinogram $Y_{para}(t, \theta)$ using the fan-to-parallel beam conversion. Then, we can reconstruct the CT image $X(u, v)$ using the Ram-Lak filtering and backprojection. The parallel-beam conversion is implemented according to the following relation
\begin{align}
    \theta &= \gamma + \beta \label{eq:conversion_1}, \\
    t &= D \sin{\gamma} \label{eq:conversion_2},
\end{align}
where $t$ is the projection location on the parallel beam detector and $\theta$ is the projection angle in the parallel-beam geometry as shown in Figure~\ref{fig:para}. For efficiency, we implement the change of variable using two 1D interpolations, one for $\theta$ in (\ref{eq:conversion_1}) and the other for $t$ in (\ref{eq:conversion_2}).

The Ram-Lak filtering for $Y_{para}(t, \theta)$ is implemented by
\begin{equation}
    Q(t, \theta) = \mathcal{F}^{-1}_t \left\{ |\omega| \cdot \mathcal{F}_t\left\{Y_{para}(t, \theta)\right\}  \right\},
\end{equation}
where $\mathcal{F}_t$ and $\mathcal{F}_t^{-1}$ are the Discrete Fourier Transform (DFT) and inverse Discrete Fourier Transform (iDFT) with respect to the detector dimension. The filtering module is implemented using the operations \texttt{torch.fft} and \texttt{torch.ifft} in PyTorch.

The backprojection module takes the filtered projection $Q(t, \theta)$ as input, and reconstructs $X(u, v)$ via
\begin{align}
     X(u, v) &= \int_0^{\pi} Q(u \cos \theta + v \sin \theta, \theta) d\theta \nonumber\\
     &\approx  \Delta \theta \sum_{i} Q(u \cos \theta_i + v \sin \theta_i, \theta_i) \nonumber\\
     &\approx \Delta \theta \sum_i (\lceil t_i \rceil - t_i)Q(\lfloor t_i \rfloor, \theta_i) \nonumber \\
     & \quad \quad + (t_i - \lfloor t_i \rfloor)Q(\lceil t_i \rceil, \theta_i),
     \label{eq:bp}
\end{align}
 where $t_i = u \cos \theta_i + v \sin \theta_i$ is a function of $u$, $v$, and $i$. The forward-pass of~\eqref{eq:bp} is parallelizable in $\theta_i$. During back-propagation, the gradients of the CT image with respect to the sinogram are given by
 \begin{equation}
     \dfrac{\partial X(u, v)}{\partial Q(t, \theta)} =
     \begin{cases}
     \Delta\theta (\lceil t_i \rceil - t_i), &\text{if} ~~ t = \lfloor t_i \rfloor,\\
         \Delta\theta (t_i - \lfloor t_i \rfloor), &\text{if} ~~ t = \lceil t_i \rceil,\\
         0, &\text{otherwise. }
     \end{cases}
 \end{equation}
 The backprojection module is implemented as a CUDA extension of PyTorch.

 \section{Evaluation on CT Images with Real Metal Artifact}
Evaluating MAR methods on CT images of patients carrying metal implants is challenging for two reasons: 1) Modern clinical CT machines have certain build-in MAR algorithms. Evaluations on CT images after MAR would not be meaningful; 2) Sinogram data with metal artifacts are difficult to access, except perhaps from machine manufacturers. To the best of our knowledge, there is no existing sinogram database which targets MAR.

In order to compare different MAR methods, we manually collect CT images with metal artifact from DeepLesion~\cite{Yan-18-lesion} and apply the following steps to obtain the metal trace $\mathcal{M}_t$ and the LI sinogram $Y_{LI}$. DuDoNet can be applied by taking $\mathcal{M}_t$ and $Y_{LI}$ as inputs. Conceptually, the steps can be understood as projecting the input CT image with unknown imaging geometry to the source domain\footnote{The domain of CT images with simulated metal artifacts.} with known geometry.

(i) $\mathcal{M}_t$: We first segment out the metal mask by applying a threshold of 2,000 HU to the metal-corrupted CT image. $\mathcal{M}_t$ can be obtained by forward projection with the imaging geometry presented in Section 4 in the manuscript.

(ii) $Y_{LI}$: We adopt the same simulation procedures and imaging geometry as in the manuscript to synthesize metal-corrupted sinogram $Y$. $Y_{LI}$ can be generated from $Y$ and $\mathcal{M}_t$ by linear interpolation.

\begin{figure}[t!]
    \vspace{-0.1cm}
    \setlength{\abovecaptionskip}{0.1pt}
    \setlength{\tabcolsep}{1pt}
    \begin{tabular}{ccc}
        \begin{subfigure}[b]{0.32\linewidth}
            \setlength{\abovecaptionskip}{1pt}
            \includegraphics[width=\textwidth,height=0.82\textwidth]{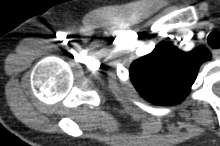}
            \caption*{Raw CT}
        \end{subfigure} &
        \begin{subfigure}[b]{0.32\linewidth}
            \setlength{\abovecaptionskip}{1pt}
            \includegraphics[width=\textwidth,height=0.82\textwidth]{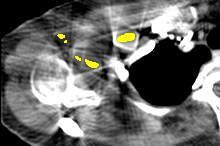}
            \caption*{LI}
        \end{subfigure} &
        \begin{subfigure}[b]{0.32\linewidth}
            \setlength{\abovecaptionskip}{1pt}
            \includegraphics[width=\textwidth,height=0.82\textwidth]{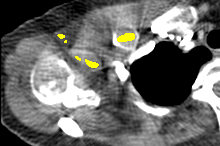}
            \caption*{NMAR}
        \end{subfigure} \\
        \begin{subfigure}[b]{0.32\linewidth}
            \setlength{\abovecaptionskip}{1pt}
            \includegraphics[width=\textwidth,height=0.82\textwidth]{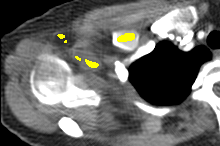}
            \caption*{RDN-CT}
        \end{subfigure} &
        \begin{subfigure}[b]{0.32\linewidth}
            \setlength{\abovecaptionskip}{1pt}
            \includegraphics[width=\textwidth,height=0.82\textwidth]{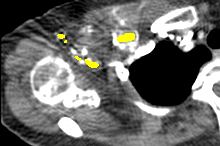}
            \caption*{CNNMAR}
        \end{subfigure} &
        \begin{subfigure}[b]{0.32\linewidth}
            \setlength{\abovecaptionskip}{1pt}
            \includegraphics[width=\textwidth,height=0.82\textwidth]{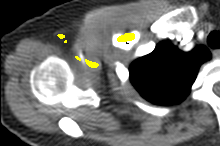}
            \caption*{DuDoNet}
        \end{subfigure}\\
        \hline
        \begin{subfigure}[b]{0.32\linewidth}
            \setlength{\abovecaptionskip}{1pt}
            \vspace{0.1cm}
            \includegraphics[width=\textwidth,height=0.82\textwidth]{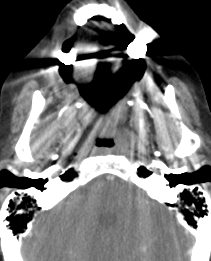}
            \caption*{Raw CT}
        \end{subfigure} &
        \begin{subfigure}[b]{0.32\linewidth}
            \setlength{\abovecaptionskip}{1pt}
            \includegraphics[width=\textwidth,height=0.82\textwidth]{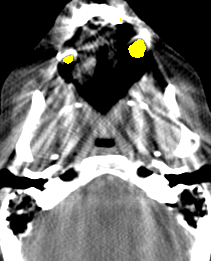}
            \caption*{LI}
        \end{subfigure} &
        \begin{subfigure}[b]{0.32\linewidth}
            \setlength{\abovecaptionskip}{1pt}
            \includegraphics[width=\textwidth,height=0.82\textwidth]{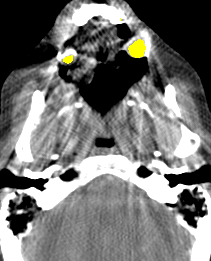}
            \caption*{NMAR}
        \end{subfigure} \\
        \begin{subfigure}[b]{0.32\linewidth}
            \setlength{\abovecaptionskip}{1pt}
            \includegraphics[width=\textwidth,height=0.82\textwidth]{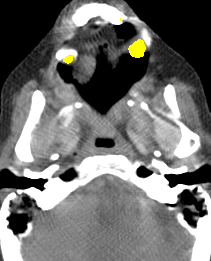}
            \caption*{RDN-CT}
        \end{subfigure} &
        \begin{subfigure}[b]{0.32\linewidth}
            \setlength{\abovecaptionskip}{1pt}
            \includegraphics[width=\textwidth,height=0.82\textwidth]{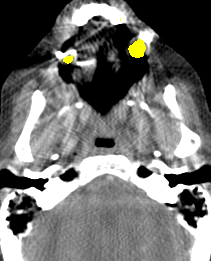}
            \caption*{CNNMAR}
        \end{subfigure} &
        \begin{subfigure}[b]{0.32\linewidth}
            \setlength{\abovecaptionskip}{1pt}
            \includegraphics[width=\textwidth,height=0.82\textwidth]{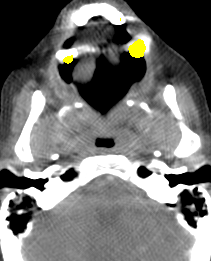}
            \caption*{DuDoNet}
        \end{subfigure}
    \end{tabular}
    \caption{Evaluations on real data. All models are exactly the same as in the main paper (no re-training).}
    \label{fig:clinical}
\end{figure}

Figure~\ref{fig:clinical} presents visual comparisons of different MAR algorithms. Metal masks obtained by step (i) are colored in yellow. 
We would like to emphasize that the true sinogram of a given CT image cannot be inferred without information about the actual imaging geometry (e.g. source to detector distance, and number of projection views). Therefore, in Figure~\ref{fig:clinical}, due to inconsistent imaging geometry, sinogram-based MAR approaches (e.g. LI) may lead to an even worse visual quality than raw CT. In contrast, DuDoNet effectively reduces metal artifacts in real CT images.

\section{Additional Visual Comparisons on CT images with Synthesized Metal Artifact}
See pages \pageref{fig:supply_visual_1} and \pageref{fig:supply_visual_2}.
\begin{figure*}[t!]
    \setlength{\abovecaptionskip}{3pt}
    \setlength{\tabcolsep}{2pt}
    \begin{tabular}[b]{cccc}
        \begin{subfigure}[b]{.24\linewidth}
            \setlength{\abovecaptionskip}{3pt}
            \includegraphics[width=\textwidth,height=0.75\textwidth]{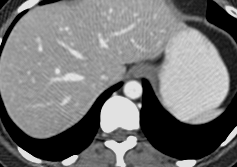}
            \caption*{Ground Truth}
        \end{subfigure} &
        \begin{subfigure}[b]{.24\linewidth}
            \setlength{\abovecaptionskip}{3pt}
            \includegraphics[width=\textwidth,height=0.75\textwidth]{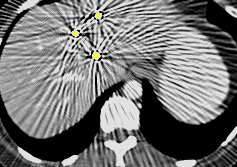}
            \caption*{With Metal Artifact}
        \end{subfigure} &
        \begin{subfigure}[b]{.24\linewidth}
            \setlength{\abovecaptionskip}{3pt}
            \includegraphics[width=\textwidth,height=0.75\textwidth]{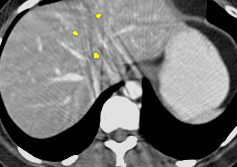}
            \caption*{LI~\cite{Kalender-87-LI}}
        \end{subfigure} &
        \begin{subfigure}[b]{.24\linewidth}
            \setlength{\abovecaptionskip}{3pt}
            \includegraphics[width=\textwidth,height=0.75\textwidth]{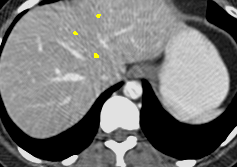}
            \caption*{NMAR~\cite{Meyer-10-nmar}}
        \end{subfigure} \\
        \begin{subfigure}[b]{.24\linewidth}
            \setlength{\abovecaptionskip}{3pt}
            \includegraphics[width=\textwidth,height=0.75\textwidth]{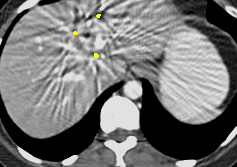}
            \caption*{cGAN-CT~\cite{Wang-18-MAR-MICCAI}}
        \end{subfigure} &
        \begin{subfigure}[b]{.24\linewidth}
            \setlength{\abovecaptionskip}{3pt}
            \includegraphics[width=\textwidth,height=0.75\textwidth]{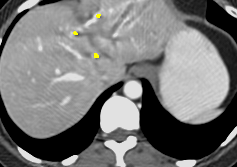}
            \caption*{RDN-CT~\cite{zhang-18-RDN}}
        \end{subfigure} &
        \begin{subfigure}[b]{.24\linewidth}
            \setlength{\abovecaptionskip}{3pt}
            \includegraphics[width=\textwidth,height=0.75\textwidth]{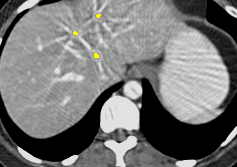}
            \caption*{CNNMAR~\cite{Zhang-18-MAR-TMI}}
        \end{subfigure} &
        \begin{subfigure}[b]{.24\linewidth}
            \setlength{\abovecaptionskip}{3pt}
            \includegraphics[width=\textwidth,height=0.75\textwidth]{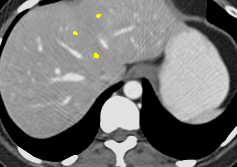}
            \caption*{DuDoNet}
        \end{subfigure} \\ \midrule \vspace{-0.5em} \\
        \begin{subfigure}[b]{.24\linewidth}
            \setlength{\abovecaptionskip}{3pt}
            \includegraphics[width=\textwidth,height=0.75\textwidth]{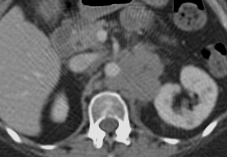}
            \caption*{Ground Truth}
        \end{subfigure} &
        \begin{subfigure}[b]{.24\linewidth}
            \setlength{\abovecaptionskip}{3pt}
            \includegraphics[width=\textwidth,height=0.75\textwidth]{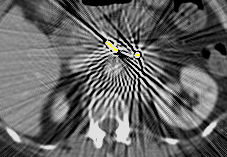}
            \caption*{With Metal Artifact}
        \end{subfigure} &
        \begin{subfigure}[b]{.24\linewidth}
            \setlength{\abovecaptionskip}{3pt}
            \includegraphics[width=\textwidth,height=0.75\textwidth]{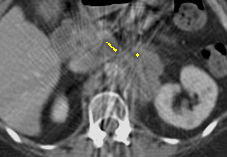}
            \caption*{LI~\cite{Kalender-87-LI}}
        \end{subfigure} &
        \begin{subfigure}[b]{.24\linewidth}
            \setlength{\abovecaptionskip}{3pt}
            \includegraphics[width=\textwidth,height=0.75\textwidth]{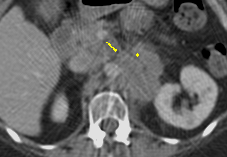}
            \caption*{NMAR~\cite{Meyer-10-nmar}}
        \end{subfigure} \\
        \begin{subfigure}[b]{.24\linewidth}
            \setlength{\abovecaptionskip}{3pt}
            \includegraphics[width=\textwidth,height=0.75\textwidth]{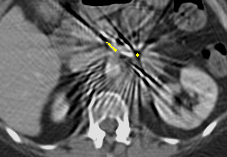}
            \caption*{cGAN-CT~\cite{Wang-18-MAR-MICCAI}}
        \end{subfigure} &
        \begin{subfigure}[b]{.24\linewidth}
            \setlength{\abovecaptionskip}{3pt}
            \includegraphics[width=\textwidth,height=0.75\textwidth]{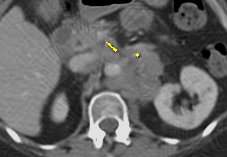}
            \caption*{RDN-CT~\cite{zhang-18-RDN}}
        \end{subfigure} &
        \begin{subfigure}[b]{.24\linewidth}
            \setlength{\abovecaptionskip}{3pt}
            \includegraphics[width=\textwidth,height=0.75\textwidth]{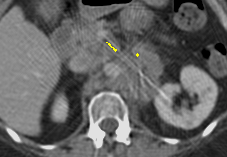}
            \caption*{CNNMAR~\cite{Zhang-18-MAR-TMI}}
        \end{subfigure} &
        \begin{subfigure}[b]{.24\linewidth}
            \setlength{\abovecaptionskip}{3pt}
            \includegraphics[width=\textwidth,height=0.75\textwidth]{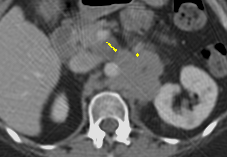}
            \caption*{DuDoNet}
        \end{subfigure} \\ \midrule \vspace{-0.5em} \\
        \begin{subfigure}[b]{.24\linewidth}
            \setlength{\abovecaptionskip}{3pt}
            \includegraphics[width=\textwidth,height=0.75\textwidth]{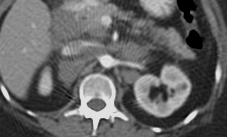}
            \caption*{Ground Truth}
        \end{subfigure} &
        \begin{subfigure}[b]{.24\linewidth}
            \setlength{\abovecaptionskip}{3pt}
            \includegraphics[width=\textwidth,height=0.75\textwidth]{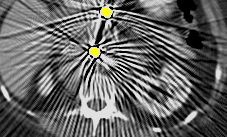}
            \caption*{With Metal Artifact}
        \end{subfigure} &
        \begin{subfigure}[b]{.24\linewidth}
            \setlength{\abovecaptionskip}{3pt}
            \includegraphics[width=\textwidth,height=0.75\textwidth]{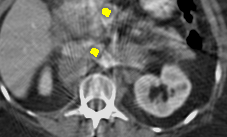}
            \caption*{LI~\cite{Kalender-87-LI}}
        \end{subfigure} &
        \begin{subfigure}[b]{.24\linewidth}
            \setlength{\abovecaptionskip}{3pt}
            \includegraphics[width=\textwidth,height=0.75\textwidth]{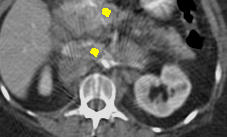}
            \caption*{NMAR~\cite{Meyer-10-nmar}}
        \end{subfigure} \\
        \begin{subfigure}[b]{.24\linewidth}
            \setlength{\abovecaptionskip}{3pt}
            \includegraphics[width=\textwidth,height=0.75\textwidth]{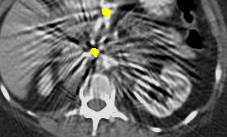}
            \caption*{cGAN-CT~\cite{Wang-18-MAR-MICCAI}}
        \end{subfigure} &
        \begin{subfigure}[b]{.24\linewidth}
            \setlength{\abovecaptionskip}{3pt}
            \includegraphics[width=\textwidth,height=0.75\textwidth]{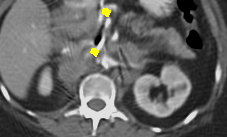}
            \caption*{RDN-CT~\cite{zhang-18-RDN}}
        \end{subfigure} &
        \begin{subfigure}[b]{.24\linewidth}
            \setlength{\abovecaptionskip}{3pt}
            \includegraphics[width=\textwidth,height=0.75\textwidth]{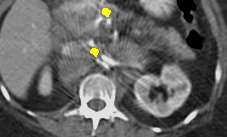}
            \caption*{CNNMAR~\cite{Zhang-18-MAR-TMI}}
        \end{subfigure} &
        \begin{subfigure}[b]{.24\linewidth}
            \setlength{\abovecaptionskip}{3pt}
            \includegraphics[width=\textwidth,height=0.75\textwidth]{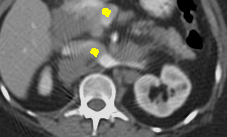}
            \caption*{DuDoNet}
        \end{subfigure} \\ 
    \end{tabular}
    \label{fig:supply_visual_1}
\end{figure*}

\begin{figure*}
    \setlength{\abovecaptionskip}{3pt}
    \setlength{\tabcolsep}{2pt}
    \begin{tabular}[b]{cccc}
        \begin{subfigure}[b]{.24\linewidth}
            \setlength{\abovecaptionskip}{3pt}
            \includegraphics[width=\textwidth,height=0.75\textwidth]{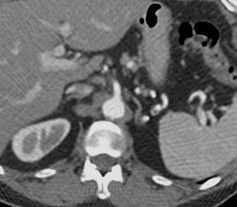}
            \caption*{Ground Truth}
        \end{subfigure} &
        \begin{subfigure}[b]{.24\linewidth}
            \setlength{\abovecaptionskip}{3pt}
            \includegraphics[width=\textwidth,height=0.75\textwidth]{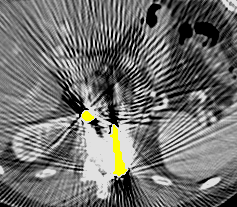}
            \caption*{With Metal Artifact}
        \end{subfigure} &
        \begin{subfigure}[b]{.24\linewidth}
            \setlength{\abovecaptionskip}{3pt}
            \includegraphics[width=\textwidth,height=0.75\textwidth]{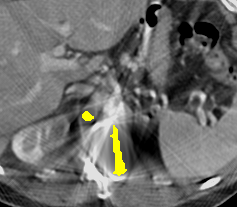}
            \caption*{LI~\cite{Kalender-87-LI}}
        \end{subfigure} &
        \begin{subfigure}[b]{.24\linewidth}
            \setlength{\abovecaptionskip}{3pt}
            \includegraphics[width=\textwidth,height=0.75\textwidth]{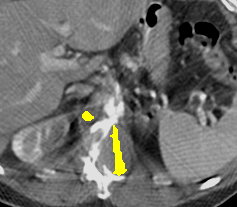}
            \caption*{NMAR~\cite{Meyer-10-nmar}}
        \end{subfigure} \\
        \begin{subfigure}[b]{.24\linewidth}
            \setlength{\abovecaptionskip}{3pt}
            \includegraphics[width=\textwidth,height=0.75\textwidth]{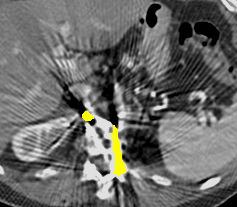}
            \caption*{cGAN-CT~\cite{Wang-18-MAR-MICCAI}}
        \end{subfigure} &
        \begin{subfigure}[b]{.24\linewidth}
            \setlength{\abovecaptionskip}{3pt}
            \includegraphics[width=\textwidth,height=0.75\textwidth]{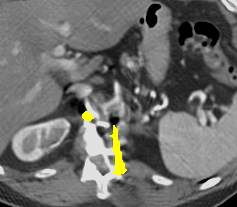}
            \caption*{RDN-CT~\cite{zhang-18-RDN}}
        \end{subfigure} &
        \begin{subfigure}[b]{.24\linewidth}
            \setlength{\abovecaptionskip}{3pt}
            \includegraphics[width=\textwidth,height=0.75\textwidth]{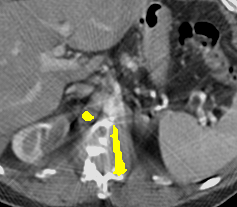}
            \caption*{CNNMAR~\cite{Zhang-18-MAR-TMI}}
        \end{subfigure} &
        \begin{subfigure}[b]{.24\linewidth}
            \setlength{\abovecaptionskip}{3pt}
            \includegraphics[width=\textwidth,height=0.75\textwidth]{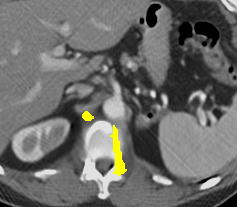}
            \caption*{DuDoNet}
        \end{subfigure} \\ \midrule \vspace{-0.5em} \\
        \begin{subfigure}[b]{.24\linewidth}
            \setlength{\abovecaptionskip}{3pt}
            \includegraphics[width=\textwidth,height=0.75\textwidth]{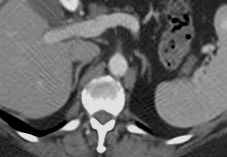}
            \caption*{Ground Truth}
        \end{subfigure} &
        \begin{subfigure}[b]{.24\linewidth}
            \setlength{\abovecaptionskip}{3pt}
            \includegraphics[width=\textwidth,height=0.75\textwidth]{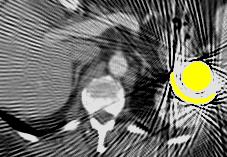}
            \caption*{With Metal Artifact}
        \end{subfigure} &
        \begin{subfigure}[b]{.24\linewidth}
            \setlength{\abovecaptionskip}{3pt}
            \includegraphics[width=\textwidth,height=0.75\textwidth]{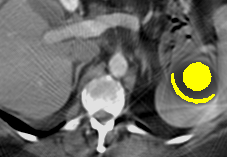}
            \caption*{LI~\cite{Kalender-87-LI}}
        \end{subfigure} &
        \begin{subfigure}[b]{.24\linewidth}
            \setlength{\abovecaptionskip}{3pt}
            \includegraphics[width=\textwidth,height=0.75\textwidth]{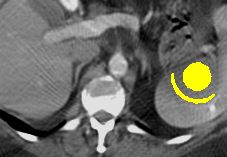}
            \caption*{NMAR~\cite{Meyer-10-nmar}}
        \end{subfigure} \\
        \begin{subfigure}[b]{.24\linewidth}
            \setlength{\abovecaptionskip}{3pt}
            \includegraphics[width=\textwidth,height=0.75\textwidth]{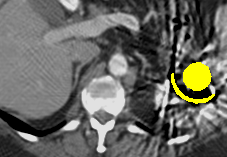}
            \caption*{cGAN-CT~\cite{Wang-18-MAR-MICCAI}}
        \end{subfigure} &
        \begin{subfigure}[b]{.24\linewidth}
            \setlength{\abovecaptionskip}{3pt}
            \includegraphics[width=\textwidth,height=0.75\textwidth]{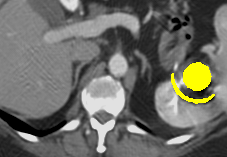}
            \caption*{RDN-CT~\cite{zhang-18-RDN}}
        \end{subfigure} &
        \begin{subfigure}[b]{.24\linewidth}
            \setlength{\abovecaptionskip}{3pt}
            \includegraphics[width=\textwidth,height=0.75\textwidth]{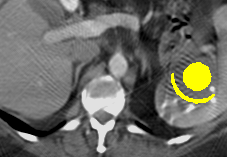}
            \caption*{CNNMAR~\cite{Zhang-18-MAR-TMI}}
        \end{subfigure} &
        \begin{subfigure}[b]{.24\linewidth}
            \setlength{\abovecaptionskip}{3pt}
            \includegraphics[width=\textwidth,height=0.75\textwidth]{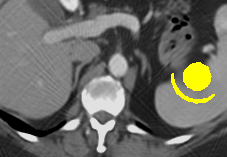}
            \caption*{DuDoNet}
        \end{subfigure} \\ \midrule \vspace{-0.5em} \\
        \begin{subfigure}[b]{.24\linewidth}
            \setlength{\abovecaptionskip}{3pt}
            \includegraphics[width=\textwidth,height=0.75\textwidth]{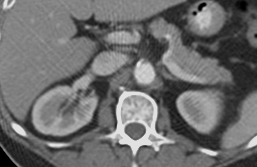}
            \caption*{Ground Truth}
        \end{subfigure} &
        \begin{subfigure}[b]{.24\linewidth}
            \setlength{\abovecaptionskip}{3pt}
            \includegraphics[width=\textwidth,height=0.75\textwidth]{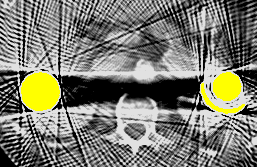}
            \caption*{With Metal Artifact}
        \end{subfigure} &
        \begin{subfigure}[b]{.24\linewidth}
            \setlength{\abovecaptionskip}{3pt}
            \includegraphics[width=\textwidth,height=0.75\textwidth]{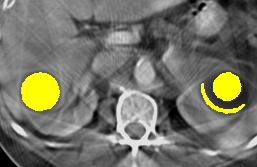}
            \caption*{LI~\cite{Kalender-87-LI}}
        \end{subfigure} &
        \begin{subfigure}[b]{.24\linewidth}
            \setlength{\abovecaptionskip}{3pt}
            \includegraphics[width=\textwidth,height=0.75\textwidth]{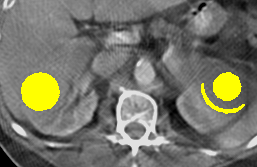}
            \caption*{NMAR~\cite{Meyer-10-nmar}}
        \end{subfigure} \\
        \begin{subfigure}[b]{.24\linewidth}
            \setlength{\abovecaptionskip}{3pt}
            \includegraphics[width=\textwidth,height=0.75\textwidth]{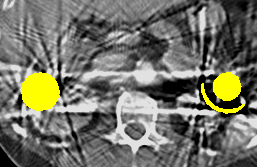}
            \caption*{cGAN-CT~\cite{Wang-18-MAR-MICCAI}}
        \end{subfigure} &
        \begin{subfigure}[b]{.24\linewidth}
            \setlength{\abovecaptionskip}{3pt}
            \includegraphics[width=\textwidth,height=0.75\textwidth]{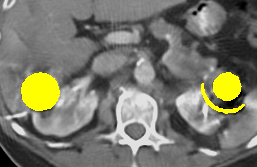}
            \caption*{RDN-CT~\cite{zhang-18-RDN}}
        \end{subfigure} &
        \begin{subfigure}[b]{.24\linewidth}
            \setlength{\abovecaptionskip}{3pt}
            \includegraphics[width=\textwidth,height=0.75\textwidth]{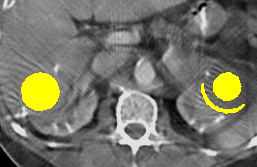}
            \caption*{CNNMAR~\cite{Zhang-18-MAR-TMI}}
        \end{subfigure} &
        \begin{subfigure}[b]{.24\linewidth}
            \setlength{\abovecaptionskip}{3pt}
            \includegraphics[width=\textwidth,height=0.75\textwidth]{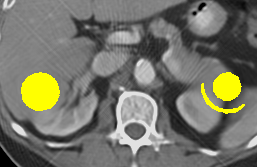}
            \caption*{DuDoNet}
        \end{subfigure} \\ 
    \end{tabular}
    \label{fig:supply_visual_2}
\end{figure*}

\section{Practical Issues}
In this section, we discuss practical issues when applying deep learning for MAR. Suppose we have access to the sinograms and CT images taken from a CT machine, methods such as CNNMAR~\cite{Zhang-18-MAR-TMI}, and cGAN-CT~\cite{Wang-18-MAR-MICCAI} require paired data, i.e., CT images with and without metallic implants from the same patient. In our approach, data within the metal trace is viewed as missing and replaced using LI~\cite{Kalender-87-LI}. Therefore, to train a DuDoNet, only implant-free sinograms, CT images and masks of metallic implants are required. In the experimental evaluations, we synthesize metal artifacts mainly for comparing with existing MAR approaches. In real applications, simulated pairs are not needed by the proposed DuDoNet.

{\small

}

\end{document}